\documentclass[preprint,10pt]{elsarticle}
\usepackage{amsmath}
\usepackage{bm}
\usepackage{adjustbox}
\usepackage{float}
\newtheorem{theorem}{Theorem}[section]
\usepackage{subcaption}
\usepackage{algorithm2e}
\newtheorem{definition}{Definition}[section]

\RestyleAlgo{ruled}
\usepackage{amssymb}
\usepackage{hyperref}
\journal{International Journal of Approximate Reasoning}

\begin{document}

\begin{frontmatter}



\title{Soft computing for the posterior  of a new matrix t graphical network}


\author[inst1,inst3]{J. Pillay}
\author[inst1,inst3]{A. Bekker}
\author[inst1,inst3]{J.T. Ferreira}
\author[inst1,inst2]{M. Arashi}

\affiliation[inst1]{organization={Department of Statistics, University of Pretoria, Pretoria, South Africa}}

\affiliation[inst2]{organization={Department of Statistics, Ferdowsi University of Mashhad, Iran}}

\affiliation[inst3]{organization={Centre of Excellence in Mathematical and Statistical Sciences, Johannesburg, South Africa}}

\begin{abstract}
Modelling noisy data in a network context remains an unavoidable obstacle; fortunately, random matrix theory may comprehensively describe network environments effectively. Thus it necessitates the probabilistic characterisation of these networks (and accompanying noisy data) using matrix variate models. Denoising network data using a Bayes approach is not common in surveyed literature. This paper adopts the Bayesian viewpoint and introduces a new matrix variate t-model in a prior sense by relying on the matrix variate gamma distribution for the noise process, following the Gaussian graphical network for the cases when the normality assumption is violated. From a statistical learning viewpoint, such a theoretical consideration indubitably benefits the real-world comprehension of structures causing noisy data with network-based attributes as part of machine learning in data science. A full structural learning procedure is provided for calculating and approximating the resulting posterior of interest to assess the considered model’s network centrality measures. Experiments with synthetic and real-world stock price data are performed not only to validate the proposed algorithm's capabilities but also to show that this model has wider flexibility than originally implied in \cite{billio2021}.

\end{abstract}

\begin{highlights}
\item An improvement on the accuracy measures, following the Bayesian estimation inference with more flexible priors, namely the matrix variate gamma and inverse matrix variate gamma as priors for  the covariance matrices. The common scale parameter $\beta$ is considered as fixed or having either a non-informative or inverse gamma distribution. Experimental results show that our proposed framework outperforms the work done by \cite{billio2021};
\item Extending the framework of denoising financial data in the context of network theory where the assumption of normality in the model is no longer sufficient to explain the variation. 
\end{highlights}


\end{frontmatter}


\section{Introduction}
\subsection{Graphical network and its reach}

Applying graph theory in real life helps understand how nodes interact; however, it is not often immediately clear which components influence others. Researchers rely on techniques to identify these hidden relationships. Robust, straightforward, and generalised methods to determine causal relationships do not exist, nor are they easy to develop. However, a graph's information is comprehensively stored in a matrix and thus is typically characterised by the adjacency matrix, whose entries denote the number of edges from one node to another. Thus, estimating a graph itself is equivalent to estimating its adjacency matrix. This realisation paves the way to exploit random matrix theory to model the graphical network. For some background on graphical models, see for instance \cite{lauritzen1996graphical}, \cite{lauritzen1989graphical} and \cite{whittaker2009graphical}.\\

Statistics and machine learning in conjunction with graph theory is a scientific area consisting of estimation of the presence of edges between graph nodes, analysing the estimation errors, and calculating the estimated graph’s structure. Evaluating the goodness-of-fit of any chosen edge estimation technique requires finding and evaluating statistics around its resulting error distribution, particularly the spread of errors about the population mean. The error’s population covariance matrix is unknown, implying that it, too, must be estimated. Frequentist estimation techniques, such as the method of moments, unconditional least squares, and maximum likelihood, exist but depend on the cleanliness of observed data to fill in most of the missing picture. In other words, assuming the observations contain the underlying pattern with error-free variation. The latter assumption poses a problem in practice as the variance of real-life data includes, among others, measurement errors and errors from rounding off. Misleading estimates and centrality measures are inevitable without incorporating additional information about the data’s variation, apriori. \\

Let $G_t$ be a sequence of networks for $t=1,\dots,T$ for $T \in \mathbb{N}$. Assume that the number of nodes does not change with respect to $t$, but the number of edges can change with respect to $t$. Assume further that each of the nodes bears a stationary time series of variables that characterises said node. Using the nodes' time series, a sequence of networks $G_t$ is estimated to represent the relationships between the nodes at time $t$ via Granger causality. This also means an adjacency matrix is estimated for $G_t$ at each time index $t$, say $\bm{Y}_t$. A node $u$ that Granger causes another node $v$ at time $t$ results in a directed edge from $u$ to $v$ at time $t$ so that the $u,v^{th}$ entry of $\bm{Y}_{t}$ is non-zero. In application, it would be redundant to consider an edge that loops around a node (i.e., an edge from node $a$ to node $a$) as it would mean the node Granger causes itself, hence self-loops are excluded in this estimation procedure. A stationary time series implies that the network structure itself at time $t$ is no more than a deviation from an underlying adjacency matrix $\bm{B}$, independent of $t$. In other words, the true graphical network structure is stationary. Thus $\bm{Y}_t$ can be considered an estimator of $\bm{B}$. It is then reasonable to view $\bm{Y}_t$ as 'noisy copy' of $\bm{B}$ given by:
	\begin{align}
		\bm{Y}_t = \bm{B} + \bm{E}_t~~ \text{for }t=1,\dots,T,
		\label{model}
	\end{align}  
	$\bm{E}_t$, $n\times n$ is a random error term, independent and identically distributed for all $t=1,\dots,T$.	In the ideal case, $\bm{Y}_t, \bm{B}$, and $\bm{E}_t$ would all be matrices with discrete entries since they constitute an adjacency matrix from (\ref{model}). Network estimation techniques rely on hypothesis tests or suitable threshold values, and the likelihood of edges bases itself on probability, which is continuous and subject to additional error. Thus, a researcher must manually binarise it to resemble an adjacency matrix through a transformation rule. It is clear that the ideal case is often a pipe dream, and $\bm{E}_t$ would have to account for the errors in likelihood testing before the binarisation. That is, $\bm{E}_t$ must have continuous entries. Continuous error random variables should be centred around zero, assuming the model fitted explains the underlying pattern adequately. Alternative distributions explored in the surveyed literature focused more on skewness in univariate cases (see \cite{skwnorm} for a comprehensive exploration). Observed data may contain additional noise, but there is insufficient motivation to consider the possibility that observations trail off too sharply to one region than others. Hence, unless there is intrinsic skewness to the data (e.g. discrete errors), a symmetric spread about zero is suitable and reasonable to assume.

\subsection{Contextualising literature}
Within the global financial system, different parts of the financial world greatly rely on each other. It is demonstrated by \cite{billio2012econometric}  that network models provide an explanation of how different items are connected in the financial settings. The dynamics of crypto assets between different bitcoin exchanges, by using a network model, were explored for example by \cite{giudici2021crypto} and \cite{chen2020lead}. More recently, \cite{ahelegbey2023network} evaluated, from a sustainable finance viewpoint, a network-based machine learning model implemented in a fintech platform aiming to assign credit ratings. \\

The approach of Bayesian graphical modelling in a variety of financial settings has been investigated, for example, by \cite{giudici2016graphical}, \cite{giudici2003improving}, \cite{barigozzi2019nets} and \cite{billio2012econometric}. \cite{bianchi2019modeling} inferred that firm-level centrality does not correlate with market values and is positively linked to realized financial losses using graphical models. By applying Bayesian time-varying networks, \cite{billio2022bayesian} detected different connectivity regimes and uncovered the role of covariates in the edge-formation process for financial networks.\\

This paper focuses on Bayesian estimation for adjacency matrices, and investigates the effect of prior information on its performance illustrated through network centrality measures. Particular attention is paid to the spread of error terms after a graphical model fit and how it influences the estimated network's structure. Hence, this paper provides an enhanced Bayesian approach to estimate the noisy connectivity structure.\\

We propose a new graphical t-model that emanates from the two role players in the new onset of the proposed superior performance of modeling the noise: the matrix variate gamma and the inverted matrix variate gamma distributions. The matrix variate gamma distribution, defined by \cite{lukacs1964applications} has the Wishart distribution (which is the distribution of the sample variance covariance matrix when sampling from a multivariate normal distribution) as a nested model. This matrix variate gamma distribution received attention in the literature amongst others by \cite{kozubowski2022matrix}, \cite{edelmann2023product}, \cite{lee2017inference}, and \cite{mathai2022singular}; for damping modeling  and diffusion MRI, see \cite{adhikari2007characterization} and  \cite{reymbaut2021matrix} respectively; for accurate signal reconstruction consult \cite{ostashev2020modeling}, \cite{jian2007continuous} and \cite{caruyer2022numerical}; for image classification, see \cite{luo2018nonparametric}.  For the interested reader, we refer to the work of Iranmanesh et al \cite{iranmanesh2013inverted} and \cite{iranmanesh2022generalized} and references therein.

\subsection{Contributions and outline}
In this paper an improvement of the previously considered model by \cite{billio2021} is proposed, since the constant shape parameter of the Wishart can now assume any integer or non-integer value. We also considered a prior on this shape parameter with surprising results when applying the centrality measures.\\

The contributions of the paper are:
\begin{itemize}
\item Highlighting the \textit{flexibility} of choosing a more general prior
distribution, via accuracy of centrality measures. In particular, the
observation of the improvement on the accuracy measures, following the
Bayesian estimation inference with matrix variate gamma and inverse matrix
variate gamma as priors for covariance matrices. Experimental results show that our proposed framework outperforms the work of \cite{billio2021}.

\item Scrutinising via detailed simulation experiments through a soft computing lense, representative of
our empirical application, to better understand the best approach needed to
improve estimation accuracy.

\item Extend the framework of denoising financial data in the context of
network theory where the assumption of normality in the model is no longer
sufficient to explain the variation. While this paper focuses on denoising
financial data, it aims to extend the framework of denoising data to any
field where applicable. Therefore, this paper combines the theory of
probability, graphs, and the descriptive properties of matrices to grow the
literature surrounding denoising graphical network data.
\end{itemize}

The remainder of the paper is organized as follows: In Section 2,  the new matrix variate graphical t-model's construction is outlined. The computational algorithm is described in Section 3, followed by an illustration of the suggested methodology via an application to real-world stock price data and a simulation study. The Conclusion in Section 4 concludes the paper.

\section{Bayesian Structure Learning of the T-Network}
There exists an extensive literature on a multivariate Gaussian distribution of errors. Articles that date back as early as the classical linear models \cite{ne} to relatively recent ones on engineering processes \cite{pe} provide an idea of how long researchers investigated the implications of this assumption. However, normality is useful when errors result from complete unexplainable variation. That is, only randomness is left over after a model's fit. However, data may also exhibit excess kurtosis (or eigenkurtosis), which implies that the observations have a large spread from their location. These factors suggest that the probabilistic nature of errors is symmetric but has excess kurtosis compared to normality \cite{ks}. Considering this additional error, the distribution of error terms is expected to be symmetric and have heavier tails. The latter is a meaningful and practical characteristic not captured by the normal distribution, and motivates the practical advantages which would be inherited from the theoretical departure of this research.\\
 
 A t-distribution thus seems a suitable choice to characterise error. In other words, hereinafter, $\bm{E}_t$ in (\ref{model}) has a matrix variate t-distribution,\\
 $\bm{E}_t \sim t_{n,n}(\bm{0},\bm{\Sigma}_1,\bm{\Sigma}_2)$,  with corresponding probability density function (pdf) \cite{guptanagar}:
		\begin{align}
		\label{errort}
		f(\bm{E}_t) &= \frac{\Gamma_n(\frac{\nu + 2n-1}{2}) }{\pi^{(\frac{n^2}{2})} \Gamma_n(\frac{\nu + n -1}{2}) } |\bm{\Sigma}_1|^{-\frac{n}{2}}|\bm{\Sigma}_2|^{-\frac{n}{2}}|\bm{I}_{n} +\bm{\Sigma}_{1}^{-1}\bm{E}_{t}\bm{\Sigma}_{2}^{-1}\bm{E}_{t}'|^{-\frac{\nu + 2n -1}{2}}, &
	\end{align}
	where $\bm{\Sigma}_1,\bm{\Sigma}_2$ are positive definite matrices, denoted as $\bm{\Sigma}_1,\bm{\Sigma}_2 >\bm{0}$. Throughout, $\boldsymbol{I}$ and $\bm{0}$ denote the identity matrix and matrix of zeros respectively with size determined by the context, and $\left\vert {}\right\vert $ denotes the determinant. $\Gamma_p(.)$ is the multivariate gamma function defined as:
 \begin{align}
		\Gamma_p (z) = \pi^{\frac{p(p-1)}{4}}\prod_{i=1}^{p}\Gamma\left(z-\frac{1}{2}(i-1)\right)
	\end{align}
	for $p$ a natural number and $\Gamma $ \ is Euler's gamma function. 
 
    By the linearity property of a matrix-variate t random matrix and since $\bm{B}$ is a constant matrix, equation (\ref{model}) implies that $\bm{Y}_t \sim t_{n,n}(\nu,\bm{B},\bm{\Sigma}_1, \bm{\Sigma}_2)$ and consequently called the matrix variate t-model.
    
\subsection{Prior specification}
    Since $\bm{B}, \bm{\Sigma}_{1}, \bm{\Sigma}_{2}$ and $\nu$ are unknown, they need to be estimated. Making use of Bayes' theorem, a posterior distribution of $\bm{B}, \bm{\Sigma}_{1}, \bm{\Sigma}_{2}$, and $\nu$ is derived, from which a sample can be simulated to estimate $\bm{B}$ using the method of moments estimator (mme). Recall that model (\ref{model}) must account for error robustly, i.e., the spread of its error terms. Thus, it is reasonable to assume a Gaussian prior for $\bm{B}$, while $\bm{\Sigma}_1$ and $\bm{\Sigma}_2$ are assessed in greater detail. That is,  $\bm{B} \sim N(\bm{0},\bm{\Omega}_{1},\bm{\Omega}_{2} )$ with corresponding pdf \cite{guptanagar}:
	\begin{align}
		\label{norm}
		&f(\bm{B}) = 
		\frac{|\bm{\Omega}_{1}|^{-\frac{n}{2}}|\bm{\Omega}_{2}|^{-\frac{n}{2}}}{(2\pi) ^{\frac{(n^2)}{2}}}\mathrm{etr} \left(-\frac{1}{2} \left( \bm{\Omega}_{2}^{-1} \left(\bm{B}\right)'\bm{\Omega}_{1}^{-1} \left(\bm{B}\right) \right) \right),&\\
        &\bm{\Omega}_{1},\bm{\Omega}_{2} >\bm{0} &\nonumber
	\end{align}
    where $etr(\cdot)$ denotes the exponential trace of a matrix. \\
    
    Since the expected value of $\bm{Y}_t$, namely, $\mathbb{E}(\bm{Y}_t)$ has to be estimated and is only defined if $\nu>1$, and it is less common for degrees of freedom to be large, assume that $\nu$ follows a truncated gamma distribution, i.e., $\nu \sim \mathcal{TG}(a_{\nu},b_{\nu},1)$, with corresponding pdf:
	\begin{align}
		f(\nu) = \left(\frac{1}{1 - F_{\nu}(1)}\right)\frac{1}{a_{\nu}^{b_{\nu}} \Gamma(a_{\nu})} \nu^{a_{\nu}-1}e^{(-\frac{\nu}{b_{\nu}})},~~ \text{for } \nu>1\text{ and 0 otherwise},
	\end{align}
	where $F_{\nu}(1) = P(\nu \leq 1)$, and $a_{\nu}, b_{\nu} >0$ respectively. \\
 
	As for the probability distributions for the covariance matrices $\bm{\Sigma}_{1}$ and $\bm{\Sigma}_{2}$, the literature focused on the  Wishart and/or inverse-Wishart as priors. However, more control is added by considering generalised distributions, namely the matrix variate gamma and inverted matrix variate gamma distributions. Recall estimation robustness is characterised by the error term's spread which is controlled by the covariance matrices $\bm{\Sigma}_{1}$ and $\bm{\Sigma}_{2}$. This reason motivates the need to choose prior probability distributions for $\bm{\Sigma}_{1}$ and $\bm{\Sigma}_{2}$ that provide meaningful information to (\ref{model}).\\
 
 Furthermore, a useful prior should also allow for the posterior pdf to depend on its information, rather than potentially be overpowered by the likelihood function. In other words, it is reasonable to install a weight parameter that allows for prior information to propagate to the posterior. To this end, let
	$\bm{\Sigma}_{1}|\gamma$ follow a matrix variate gamma distribution. That is, $\bm{\Sigma}_{1}|\gamma\sim \mathcal{MG}_n(\delta_1, \beta, (\gamma\bm{\Phi}_1)^{-1})$ with corresponding pdf \cite{matrixgamma}:
	\begin{align}
		\label{mgpdf}
		f(\bm{\Sigma}_1) = \frac{ |\bm{\Phi}_1^{-\delta_1}| }{ \beta_1^{n\delta_1}\Gamma_n(\delta_1) }\mathrm{etr}\left\{ -\frac{1}{\beta_1} \bm{\Phi}_1^{-1}\bm{\Sigma}_1\right\} |\bm{\Sigma}_1|^{\delta_1 - \frac{n+1}{2} } ,
	\end{align} 
	 and $\bm{\Sigma}_{2}|\gamma$ follow an inverted matrix variate gamma distribution. That is,  $\bm{\Sigma}_{2}|\gamma \sim \mathcal{IMG}_n(\delta_2, \beta, (\gamma\bm{\Phi}_2)^{-1})$ with corresponding pdf \cite{matrixgamma}:
	 \begin{align}
	 	\label{imgpdf}
	 	f(\bm{\Sigma}_2) = \frac{ |\bm{\Phi}_2^{-\delta_2}| }{ \beta_2^{n\delta_2}\Gamma_n(\delta_2) }\mathrm{etr}\left\{ -\frac{1}{\beta_2} \bm{\Phi}_2^{-1}\bm{\Sigma}_2^{-1}  \right\} |\bm{\Sigma}_2|^{\delta_2 - \frac{n+1}{2} } 
	 \end{align} 
	 where $\delta_2 > \frac{n-1}{2}$ and $\beta_2>0$.
	 The common scale $\gamma$ allows the prior dependence of the unconditional distribution of $\bm{\Sigma}_{1}, \bm{\Sigma}_{2}$ to vary. The weight parameter $\gamma$ is unknown and thus must be estimated. Hence assume $\gamma \sim \mathcal{G}(a_{\gamma},b_{\gamma})$ with corresponding pdf:
	 	\begin{align}
	 	\label{gamma}
	 	f(\gamma) = \frac{1}{b_{\gamma}^{a_{\gamma}} \Gamma(a_{\gamma})} \gamma^{a_{\gamma}-1}e^{(-\frac{\gamma}{b_{\gamma}})},
	 \end{align}
  where $a_{\gamma}, b_{\gamma} >0$ respectively.\\
  
	Subsequently, the joint prior pdf is:
	\begin{align}
		\label{priors2}
		\pi(\bm{B},\bm{\Sigma}_{1},\bm{\Sigma}_{2},\gamma, \nu)
		& \propto
		\gamma^{a_{\gamma}-1 +n(\delta_1 + \delta_2)} \nu^{a_{\nu}-1} \mathrm{etr}\left\{ -\frac{1}{2}\mathrm{vec}(\bm{B})'(\bm{\Omega}_{2}\otimes\bm{\Omega}_{1})^{-1}\mathrm{vec}(\bm{B}) \right\}&\nonumber\\
		&\times \mathrm{exp}\left\{ -\left(\frac{\gamma}{b_{\gamma}} + \frac{\nu}{b_{\nu}} \right) \right\}\beta^{-(n\delta_1 + n \delta_2)}|\bm{\Sigma}_1|^{\delta_1 - \frac{n+1}{2}  }|\bm{\Sigma}_2|^{\delta_2 - \frac{n+1}{2}  } &\nonumber\\
		&\times\mathrm{etr}\left[- \gamma \left( \frac{1}{\beta}\bm{\Phi}_1\bm{\Sigma}_{1} + \frac{1}{\beta}\bm{\Phi}_2\bm{\Sigma}_2^{-1} \right) \right].&
	\end{align}
	Let $\beta_1=\beta_2=\beta$ for $\bm{\Sigma}_{1}, \bm{\Sigma}_{2}$ to explore the effect of its added control. \\

  It's valuable to be reminded of the useful relationship between the vector variate and the corresponding matrix variate distribution:
    A random matrix $\bm{B}~~n \times n$ is said to be matrix variate normal distributed with parameters $\bm{0}~, \bm{\Omega}_{1}>\bm{0}$, and $ \bm{\Omega}_{2}>\bm{0}$ denoted as $\bm{B} \sim N_{n,n}(\bm{0}~, \bm{\Sigma}_{1},~\bm{\Sigma}_{2}) \iff \mathrm{vec}(\bm{B}) \sim N_{nn}(\mathrm{vec}(\bm{0}), \bm{\Omega}_{2} \otimes\bm{\Omega}_{1}) $, where $\otimes $ stands for the Kronecker product. That is, $\mathrm{vec}(\bm{B})$  is a parsimonious reparameterisation of $\bm{B}$ that still contains all the distributional properties of its matrix predecessor. Thus, where possible, $\mathrm{vec}(\bm{B})$ is used.
    \subsection{Posterior approximation}
         Denote the random sample of estimated adjacency matrices and the collection of unknown and known parameters, respectively as follows:
     \begin{align}
         \bm{Y} &= (\bm{Y}_1,\dots,\bm{Y}_\text{T})&\\
         \bm{\theta} &=(\bm{B}, \bm{\Sigma}_{1},\bm{\Sigma}_{2},\gamma,\nu)&\\
         \bm{\phi} &= (\bm{\Omega}_1,\bm{\Omega}_2,\bm{\Psi}_1,\bm{\Psi}_2,a_{\gamma},b_{\gamma},a_{\nu},b_{\nu}).&
     \end{align}
     Then $\bm{\theta}$ has the likelihood function:
	\begin{align}
		\label{likelihood}
		\prod_{t=1}^\text{T} \frac{\Gamma_n(\frac{\nu + 2n-1}{2}) }{\pi^{(\frac{n^2}{2})} \Gamma_n(\frac{\nu + n -1}{2}) } |\bm{\Sigma}_1|^{-\frac{n}{2}}|\bm{\Sigma}_2|^{-\frac{n}{2}}|\bm{I}_{n}+\bm{\Sigma}_{1}^{-1}(\bm{Y}_{t} - \bm{B})\bm{\Sigma}_{2}^{-1}(\bm{Y}_{t} - \bm{B})'|^{-\frac{\nu + 2n -1}{2}}.
	\end{align}
	Now that the prior pdf and the likelihood function are established, by equations (\ref{priors2}) and (\ref{likelihood}), the posterior pdf $f(\bm{\theta}| \bm{Y}, \bm{\phi}) $ is proportional to:
	\begin{align}
		\label{post}
		&\prod_{t=1}^\text{T} \frac{\Gamma_n(\frac{\nu + 2n-1}{2}) }{\pi^{(\frac{n^2}{2})} \Gamma_n(\frac{\nu + n -1}{2}) } |\bm{\Sigma}_1|^{-\frac{n}{2}}|\bm{\Sigma}_2|^{-\frac{n}{2}}|\bm{I}_{n}+\bm{\Sigma}_{1}^{-1}(\bm{Y}_{t} - \bm{B})\bm{\Sigma}_{2}^{-1}(\bm{Y}_{t} - \bm{B})'|^{-\frac{\nu + 2n -1}{2}}&\nonumber\\ 
		&\times \gamma^{a_{\gamma}-1}  \nu^{a_{\nu}-1} \mathrm{etr}\left[-\frac{1}{2}\left( \mathrm{vec}(\bm{B})'(\bm{\Omega}_{2}\otimes\bm{\Omega}_{1})^{-1}\mathrm{vec}(\bm{B}) \right)\right] \mathrm{exp}\left[-\left(\frac{\gamma}{b_{\gamma}} + \frac{\nu}{b_{\nu} } \right) \right] &\nonumber\\
		& \times \mathrm{exp}\left\{ -\left(\frac{\gamma}{b_{\gamma}} + \frac{\nu}{b_{\nu}} \right) \right\}\beta^{-(n\delta_1 + n \delta_2)}|\bm{\Sigma}_1|^{\delta_1 - \frac{n+1}{2}  }|\bm{\Sigma}_2|^{\delta_2 - \frac{n+1}{2}  }&\nonumber\\
  & \times \mathrm{etr}\left[- \gamma \left( \frac{1}{\beta}\bm{\Phi}_1\bm{\Sigma}_{1} + \frac{1}{\beta}\bm{\Phi}_2\bm{\Sigma}_2^{-1} \right) \right].&
	\end{align} 
	
	It is clear (\ref{post}) cannot be represented in closed form. Hence, pursuing an analytical expression for the expected value of $\bm{\theta}$ is not feasible. (See also \cite{liu1995ml}.) Instead, this paper follows the strategy that \cite{thompson2020classification} and  \cite{billio2021} employed, considering the data-augmented approach that approximates the likelihood function and produces closed-form full conditional posterior pdfs. 
	For the latter consider the following theorem:
	\begin{theorem}
		\label{thm}
		Let $\bm{W}_t \sim \mathcal{IMG}_n(\frac{\nu + n-1}{2},2, \bm{\Sigma}_1)$ and $\bm{Y}_t|\bm{W}_t \sim N_{n,n}(\bm{B},\bm{W}_t,\bm{\Sigma}_{2})$, then $\bm{Y}_t \sim t_{n,n}(\nu, \bm{B},\bm{\Sigma}_{1},\bm{\Sigma}_{2})$.
	\end{theorem}
 Denote the collection of random matrices $\bm{W} = (\bm{W}_1, \dots, \bm{W}_{\text{T}})$. From theorem \ref{thm}, the augmented likelihood is given as:
	\begin{align}
		\label{al}
		&\prod_{t=1}^{\text{T}}\frac{|\bm{\Sigma}_2|^{-\frac{n}{2}} |\bm{\Sigma}_1|^{\frac{\nu + n-1}{2} }} {\pi^{\frac{n^2}{2} } \left(2^{\frac{n(\nu+n-1)}{2} } \right) \Gamma_n (\frac{\nu + n-1}{2}) } \left|\bm{W}_t\right|^{-\frac{ 1 }{2} \left(\nu +3n \right) } &\nonumber\\
        & \times \mathrm{etr}\left(-\frac{1}{2}\left(\bm{W}_t^{-1} (\bm{Y}_t-\bm{B})\bm{\Sigma}_{2}^{-1}(\bm{Y}_t-\bm{B})' + \bm{W}_t^{-1}\bm{\Sigma}_1  \right) \right)&\nonumber\\
		&\propto \left[2^{\frac{n(\nu+n-1)}{2}}\Gamma_n \left(\frac{\nu + n-1}{2} \right) \right]^{-n} |\bm{\Sigma}_2|^{-\frac{\text{T}n}{2}} |\bm{\Sigma}_1|^{\frac{\text{T}(\nu + n-1)}{2} } \left(\prod_{t=1}^{\text{T}} \left|\bm{W}_t \right| \right)^{-\frac{ 1 }{2} \left(\nu +3n \right)}&\nonumber\\
		&\times \mathrm{etr}\left( -\frac{1}{2} \left( \bm{\Sigma}_2^{-1}\sum_{t=1}^{\text{T}}(\bm{Y}_t - \bm{B})'\bm{W}_t^{-1}(\bm{Y}_t - \bm{B}) + \bm{\Sigma}_1\sum_{t=1}^{\text{T}}\bm{W}_t^{-1}
		\right)
		\right).
	\end{align}
	Finally, from (\ref{al}) and (\ref{priors2}) the posterior pdf is approximated by:
	\begin{align}
		\label{aug_posterior}
		&f(\bm{B},\bm{\Sigma}_{1},\bm{\Sigma}_{2},\gamma,\nu) &\nonumber \\
        &\propto \left[2^{\frac{n(\nu+n-1)}{2}}\Gamma_n \left(\frac{\nu + n-1}{2} \right) \right]^{-n}  |\bm{\Sigma}_1|^{\frac{\text{T}(\nu + n-1)}{2} }
		|\bm{\Sigma}_2|^{-\frac{\text{T}n}{2}} \left(\prod_{t=1}^{\text{T}} \left|\bm{W}_t \right| \right)^{-\frac{ 1 }{2} \left(\nu +3n \right)}&\nonumber\\
		&\times \mathrm{etr}\left( -\frac{1}{2} \left( \bm{\Sigma}_2^{-1}\sum_{t=1}^{\text{T}}(\bm{Y}_t - \bm{B})'\bm{W}_t^{-1}(\bm{Y}_t - \bm{B}) + \bm{\Sigma}_1\sum_{t=1}^{\text{T}}\bm{W}_t^{-1}
		\right)
		\right)&\nonumber\\
		&\times \gamma^{a_{\gamma}-1}  \nu^{a_{\nu}-1} \mathrm{etr}\left[-\frac{1}{2}\left( \mathrm{vec}(\bm{B})'(\bm{\Omega}_{2}\otimes\bm{\Omega}_{1})^{-1}\mathrm{vec}(\bm{B}) \right)\right]&\nonumber\\
		&\times \mathrm{exp}\left[-\left(\frac{\gamma}{b_{\gamma}} + \frac{\nu}{b_{\nu} } \right) \right]\mathrm{exp}\left[-\left(\frac{\gamma}{b_{\gamma}} + \frac{\nu}{b_{\nu} } \right) \right] &\nonumber\\
		& \times \mathrm{exp}\left\{ -\left(\frac{\gamma}{b_{\gamma}} + \frac{\nu}{b_{\nu}} \right) \right\}\beta^{-(n\delta_1 + n \delta_2)}|\bm{\Sigma}_1|^{\delta_1 - \frac{n+1}{2}  }|\bm{\Sigma}_2|^{\delta_2 - \frac{n+1}{2}  } &\nonumber \\
        & \times \mathrm{etr}\left[- \gamma \left( \frac{1}{\beta}\bm{\Phi}_1\bm{\Sigma}_{1} + \frac{1}{\beta}\bm{\Phi}_2\bm{\Sigma}_2^{-1} \right) \right].&
	\end{align}
	$\bm{\Sigma}_1$ and $\bm{\Sigma}_2$ control the spread of $\bm{E}_t$ but do not control the kurtosis of $f(\bm{E}_t)$. However, if one of the covariance matrices instead varies, it adds variance. Furthermore, if the probabilistic variation of the covariance follows an informative distribution, it introduces 'modified copies' of $\bm{E}_t$ that increase $f(\bm{E}_t)$'s kurtosis.\\
	\\
    Note that the univariate scale parameter $\beta$ in (\ref{aug_posterior}) could either be fixed or assumed unknown and estimated by imposing a prior on it, much like $\gamma$. This paper investigates the impact of including this scale parameter $\beta$. First, it fixes a range of candidate values and then imposes a non-informative Jeffrey's prior and examines how much information the likelihood contributes to estimating $\beta$. Lastly, it assumes an inverse gamma pdf as an informative prior and its performance is compared to the non-informative scenario. \\
    \\
    An observation from the distribution with pdf (\ref{aug_posterior}) estimates the adjacency matrix $\bm{B}$ in (\ref{model}). However, the expression for the pdf (\ref{aug_posterior}) is not in closed form, but most of its full conditional distributions are, which are given below:
    \begin{flalign}
        \label{nupost}
            1.&~ f(\nu| \bm{Y}, \bm{B}, \bm{\Sigma}_1,\bm{\Sigma}_2):&\nonumber\\
            & \propto \nu ^{a_{\nu} -1} e^{-\frac{\nu}{\overline{b_{\nu}}} }\left(\frac{\Gamma_n(\frac{\nu + 2n-1}{2}) }{\Gamma_n(\frac{\nu + n -1}{2}) }\right)^{\text{T}},  \text{where } &\nonumber\\
            &\overline{ b_{\nu} }=\left( \frac{1}{b_{\nu}} + \frac{1}{2} \sum_{t=1}^{\text{T}} ln( |\bm{I}_{n} +\bm{\Sigma}_{1}^{-1}(\bm{Y}_{t} - \bm{B})\bm{\Sigma}_{2}^{-1}(\bm{Y}_{t} - \bm{B})'|) \right)^{-1}.&
        \end{flalign}
        \begin{flalign}
            2.&~f(\bm{W}_t| \bm{Y}, \bm{B}, \bm{\Sigma}_1,\bm{\Sigma}_2,\nu):&\nonumber\\
            & \propto |\bm{W}_t|^{- \frac{(\nu +2n -1 + n + 1)}{2} }\mathrm{etr}\left( -\frac{1}{2}\left( (\bm{Y}_t - \bm{B})\bm{\Sigma}_2^{-1}(\bm{Y}_t - \bm{B})' + \bm{\Sigma}_1 \right) \bm{W}_t^{-1}\right)&\nonumber\\
            &\text{which is proportional to an inverted matrix variate gamma pdf with scale matrix } &\nonumber\\
            &\bm{\overline{W}} \text{degree of freedom } p,\text{ where } \bm{\overline{W}} = (\bm{Y}_t - \bm{B})\bm{\Sigma}_2^{-1}(\bm{Y}_t - \bm{B})' + \bm{\Sigma}_1&\nonumber\\
            &\text{and } p = \frac{\nu + 2n-1}{2}, \text{ and }\beta=2.&
  	\end{flalign}
    	\begin{flalign}
            3.&~f(\mathrm{vec}(\bm{B})|\mathrm{vec}(\bm{Y}),\mathrm{vec}(\bm{W}), \bm{\Sigma}_{2}\otimes\bm{\Sigma}_{1}):&\nonumber\\
        &\propto \mathrm{exp}\left( -\frac{1}{2}\left(\mathrm{vec}(\bm{B}) - \mathrm{vec}(\bm{M})\right)' \bm{\overline{\Omega}}^{-1}\left(\mathrm{vec}(\bm{B}) - \mathrm{vec}(\bm{M})\right)' \right)&\nonumber\\
        &\text{which is proportional to a vector-variate normal pdf with mean } \mathrm{vec}(\bm{M})&\nonumber\\
        &\text{ and covariance matrix }\bm{\overline{\Omega}},\text{where } 
        \bm{\overline{\Omega}} = \left[\sum_{t=1}^{\text{T}}(\bm{\Sigma}_{2} \otimes \bm{W}_{t})^{-1} + (\bm{\Omega}_{2} \otimes \bm{\Omega}_{1})^{-1}\right]^{-1} ,&\nonumber\\
        &\text{ and } \mathrm{vec}(\bm{M}) = \bm{\overline{\Omega}}\sum_{t=1}^{\text{T}}(\bm{\Sigma}_{2} \otimes \bm{W}_{t})^{-1} \mathrm{vec}(\bm{Y}_t).&
        \end{flalign}
        \begin{flalign}
            4.&~f(\bm{\Sigma}_{1}|\bm{W},\gamma,\nu,\beta):&\nonumber\\
            &\propto\mathrm{etr}\left\{-\frac{1}{\beta}\left(\frac{\beta}{2}\sum_{t=1}^{T}\bm{W}_t^{-1}\bm{\Sigma}_1 + \gamma\bm{\Phi}_1\bm{\Sigma}_1\right) \right\}|\bm{\Sigma}_1|^{\frac{\text{T}(\nu + n-1)}{2} + \delta_1 - \frac{n+1}{2} }&\nonumber\nonumber\\
    		&\text{which is proportional to a matrix variate gamma pdf with scale  parameters }&\nonumber\\
    		& \overline{\bm{\Phi}_1},\beta,\text{ and shape parameter }\overline{\delta_1} \text{ where } \overline{\delta_1} = \delta_1 + \frac{\text{T}(\nu + n-1)}{2} & \nonumber\\
    		&\text{ and } \overline{\bm{\Phi}_1}  = \left(\frac{\beta}{2}\sum_{t=1}^{\text{T}}\bm{W}_t^{-1} + \gamma\bm{\Phi}_1\right)^{-1}.&
        \end{flalign}
        \begin{flalign}
    		5.&~f(\bm{\Sigma}_{2}|\bm{Y},\bm{W}, \bm{B},\bm{\Sigma}_{1},\gamma):&\nonumber\\
        	&\propto|\bm{\Sigma}_2|^{-(\frac{Tn + n+1}{2} + \delta_2) }\mathrm{etr}\left\{ -\frac{1}{\beta} \left(\frac{\beta}{2}\sum_{t=1}^{\text{T}}(\bm{B}-\bm{Y}_t)'\bm{W}_t^{-1}(\bm{B}-\bm{Y}_t) +\gamma\bm{\Phi}_2 \right) \bm{\Sigma}_2^{-1} \right\}&\nonumber\\
    		&\text{which is proportional to an inverted matrix variate gamma pdf with}&\nonumber\\
    		&\text{ scale parameters }\overline{\bm{\Phi}}_2, \beta \text{ and shape parameter }\overline{\delta}_2
    	\text{ where } \overline{\delta}_2 = \delta_2 + \frac{\text{T}n}{2} \text{ and } &\nonumber\\
    	&\overline{\bm{\Phi}}_2 = \left( \frac{\beta}{2}\sum_{t=1}^{\text{T}}(\bm{B}-\bm{Y}_t)'\bm{W}_t^{-1}(\bm{B}-\bm{Y}_t) +\gamma\bm{\Phi}_2 \right)^{-1}. &
   	 \end{flalign}
        \begin{flalign}
    	6.&~f(\gamma|\bm{\Sigma}_1, \bm{\Sigma}_2):&\nonumber\\
    	&\propto \gamma^{\left( n\frac{\kappa_1 + 2a_z} {2} + a_{\gamma}-1 \right)}
    	\mathrm{exp}\left(-\gamma \left( \frac{1}{b_{\gamma}} + \frac{1}{2 }\mathrm{tr}\left( \bm{\Psi}_1\bm{\Sigma}_1 + \bm{\Psi}_2\bm{\Sigma}_2^{-1}\right) \right) \right)&\nonumber\\
    	&\text{which is proportional to a gamma pdf with shape and scale}&\nonumber\\
    	&\text{ parameters }\overline{a_{\gamma}}\text{ and }\overline{b_{\gamma}} , \text{ respectively. } \text{where } \overline{a_{\gamma}} = n\frac{\kappa_1 + 2a_z} {2} + a_{\gamma} \text{ and } &\nonumber\\
    	&\overline{b_{\gamma}} = \left( \frac{1}{b_{\gamma}} + \frac{1}{2 }\mathrm{tr}\left( \bm{\Psi}_1\bm{\Sigma}_1 + \bm{\Psi}_2\bm{\Sigma}_2^{-1}\right) \right)^{-1}.&
	    \end{flalign}
     If a prior pdf is imposed on $\beta$ then the full conditional posterior pdf is given as:
     \begin{flalign}
		\label{beta3}
		7.~&f(\beta| \bm{\Sigma}_1,\bm{\Sigma}_2,\gamma) \nonumber\\
        &\propto \beta^{-(n\delta_1 +n\delta_2 )}\mathrm{etr}\left\{-\gamma\left( \bm{\Phi}_1 \bm{\Sigma}_1 + \bm{\Phi}_2\bm{\Sigma}^{-1}_2 \right) \frac{1}{\beta} \right\}\times f(\beta), &\nonumber\\
		&\text{where } f(\beta) \text{ is the chosen prior}. &
	\end{flalign}

 \section{Application and computational experiments}
 \subsection{A Gibbs algorithm}
 This section discusses the Bayesian computational algorithm through hierarchical phases. Model evaluation metrics for diagnostics and comparisons are given in the Appendix. \\

An observation from the distribution with pdf (\ref{aug_posterior}) estimates the adjacency matrix $\bm{B}$ in (\ref{model}). However, the expression for the pdf (\ref{aug_posterior}) is not in closed form, but most of its conditional distributions are. Hence, this section outlines a useful algorithm via a soft computing technique to solve the complex computational problem (\ref{aug_posterior}).  We approximate inference of the parameters via a Gibbs sampler \cite{gibbs_sampler}, the algorithm is described to simulate sample of size $n$ of a collection of observations $\{\bm{B}, \bm{W}_1,\dots,\bm{W}_{\text{T}}, \bm{\Sigma}_1,\bm{\Sigma}_2,\gamma,\nu, \beta \}$ (excluding the burn-in sample).\\

	\begin{algorithm}[H]

        \caption{Gibbs algorithm}\label{gibbs2}
            Initialise parameters, $\bm{\theta}^{(0)}= (\nu^{(0)}, \bm{W}^{(0)},\bm{B}^{(0)},\bm{\Sigma}_{1}^{(0)},\bm{\Sigma}_{2}^{(0)},\gamma^{(0)}, \beta^{(0)} $).\
            Subscript elements as follows: $\bm{\theta}^{(0)} = (\bm{\theta}^{(0)}_1 ,\bm{\theta}^{(0)}_2 ,\bm{\theta}^{(0)}_3 ,\bm{\theta}^{(0)}_4 ,\bm{\theta}^{(0)}_5 \bm{\theta}^{(0)}_6 ,\bm{\theta}^{(0)}_7)$.\
            \\
            \For{$i=1$ \KwTo $n$}
            {   \For{$j=1$ \KwTo $7$}
                {   
                 from distribution with full conditional pdfs \eqref{nupost} to \eqref{beta3} :
                \\
                Sample $\bm{\theta}^{(i)}_j \sim f(\bm{\theta}_j| \bm{\Theta}^{(i-1)},\bm{\Theta}^{(i)} ) $ , where $\bm{\Theta}^{(i-1)} = \{ \bm{\theta}^{(i-1)}_k: k>j \} $ and $\bm{\Theta}^{(i)} = \{ \bm{\theta}^{(i)}_k: k<j \} $
                }
                 Then $\bm{\theta}^{(i)}= (\bm{\theta}^{(i)}_1 ,\bm{\theta}^{(i)}_2 ,\bm{\theta}^{(i)}_3,\bm{\theta}^{(i)}_4 ,\bm{\theta}^{(i)}_5 ,\bm{\theta}^{(i)}_6 ,\bm{\theta}^{(i)}_7)$, which is an observation from pdf (\ref{aug_posterior}).
            }
        \end{algorithm}
 \subsection{Computational specifications}
The simulations were run on MATLAB R2022b on the University Stats server. Runtime for simulations was 16h excluding time to compute Granger causality test statistics. Detailed specifications are as follows:\\
  
  \begin{tabular}{ll}
  Server details &: Single power Server (SuperMicro, USA).\\
  Cores & : 2  INTEL Skylake-e Central Processors, \\
   &\hspace{0.05cm}(24 CPU-cores each).\\
    CPU Clock base speed &: 2.6 GHz. \\
    Hardware memory &: RAID-1 configured 800GB SSD.\\
    LINUX support &: SuSE, SLES, SP15.3.\\
    Workspace support &: RAID-6 to support 10TB (Spindle drives).
\end{tabular}
\subsection{Simulation study}
    In order to illustrate the proposed procedure in section 2, we performed a simulation study. The methodology uses algorithm \ref{gibbs2} to estimate $\bm{B} $ for degrees of freedom ranging from 1 to 20. To investigate the impact and role that $\beta$ plays, the methodology is applied with $\beta$ set to a fixed value.  The role of $\gamma$  was explored by \cite{billio2021}.\\
    
\begin{figure}[H]
         \centering
         \includegraphics[width=0.6\textwidth]{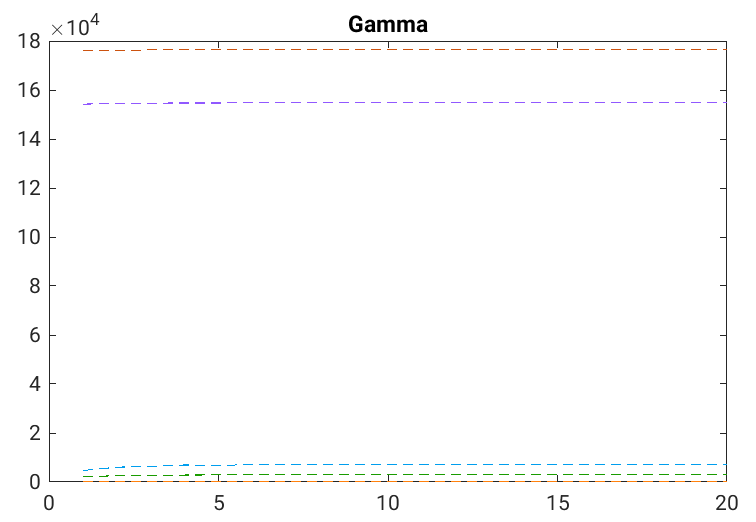}
         \caption{The dashed lines are the averages of samples obtained from algorithm \ref{gibbs2} for different fixed values of $\beta$.}
         \label{gammas}
        \end{figure}

\begin{figure}[H]
		\fbox{\parbox{\textwidth}{
				\begin{subfigure}[H]{0.5\textwidth}
					\includegraphics[width=\textwidth]{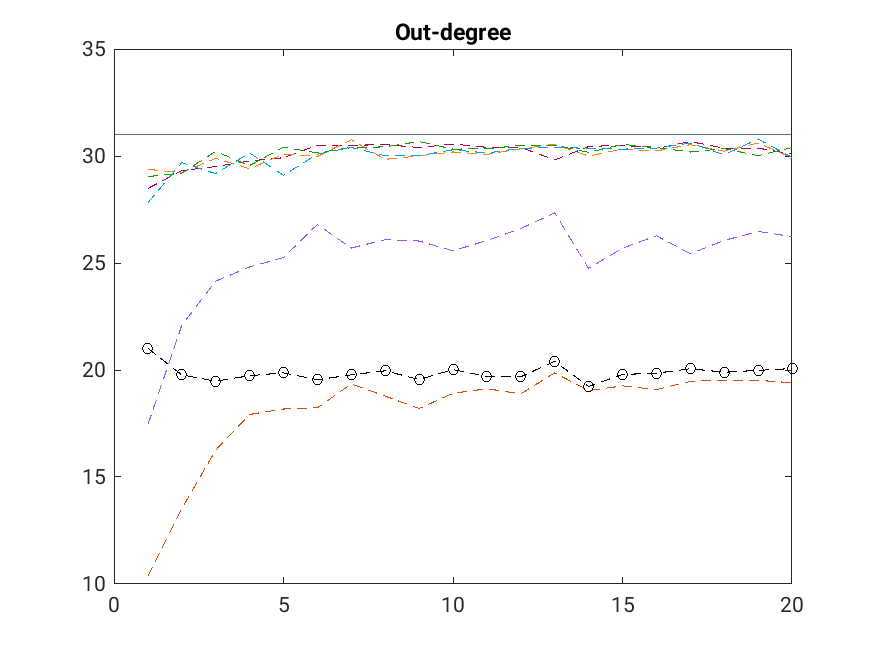}
				\end{subfigure}%
				\begin{subfigure}[H]{0.5\textwidth}
					\includegraphics[width=\textwidth]{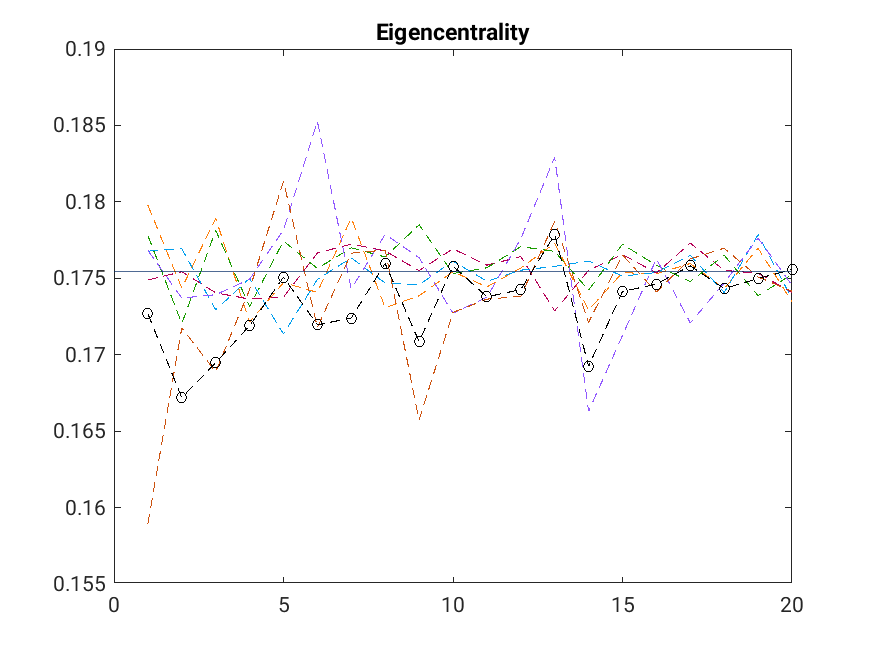}
				\end{subfigure}
                     \newline
				\begin{subfigure}[H]{0.5\textwidth}
					\includegraphics[width=\textwidth]{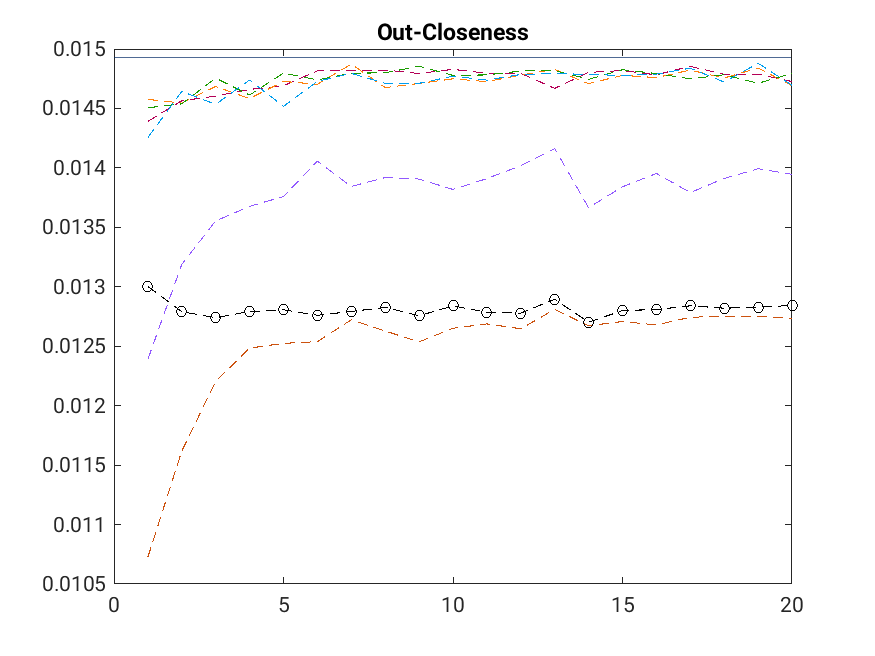}
				\end{subfigure}%
				\begin{subfigure}[H]{0.5\textwidth}
					\includegraphics[width=\textwidth]{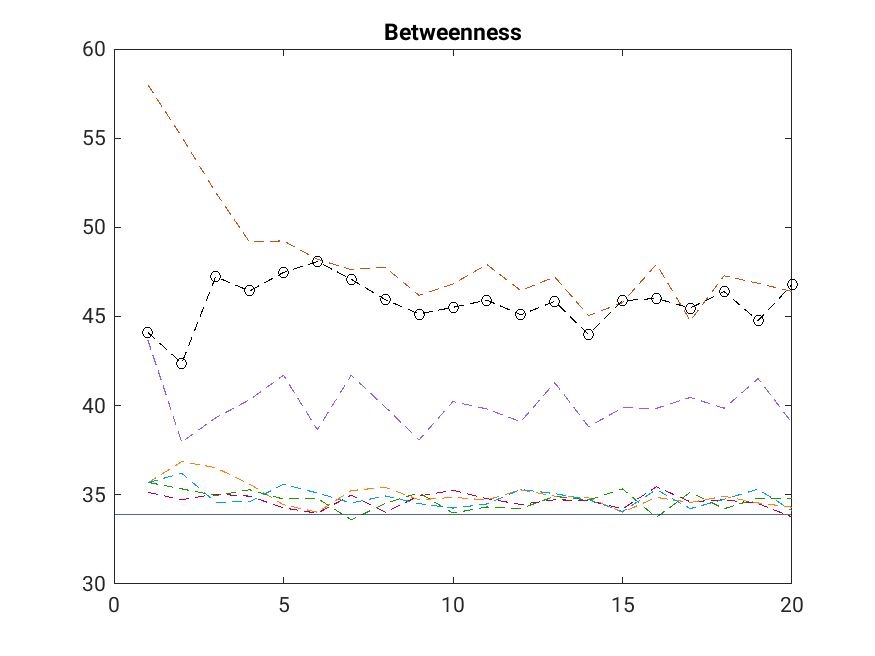}
				\end{subfigure}
			\newline
                \centering
			\begin{subfigure}[H]{0.32\textwidth}
				\includegraphics[width=\textwidth]{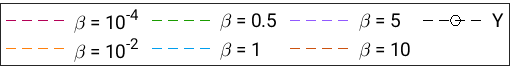}
			\end{subfigure}
		}}
  \caption{The solid grey line represents the true value, the black bubble-dashed line represents the raw data averages, and the dashed lines are the averages of samples obtained from algorithm \ref{gibbs2} for different fixed values of $\beta$.}
    \label{c2}
	\end{figure}

    The following is observed from Figures \ref{gammas} and \ref{c2}:
    \begin{itemize}
        \item[i.] The roles of $\gamma$ and $\beta$ control the extent of influence the prior and likelihood information has on the posterior pdf.
        \item[ii.] Larger values of $\beta$ cause the likelihood to have a greater influence on the posterior pdf. For instance, refer to Figure \ref{c2}. For $\beta=10$, the out-degree, out-closeness, and betweenness estimations are guided by the likelihood function almost entirely. 
        \item[iii.]Notice from Figure 1 that, despite this, the average value of $\gamma$ increases by $10^4$ to compensate for the misleading impact $\beta$ has on the estimation accuracy. In other words, there is no clear linear relationship between $\gamma$ and $\beta$.
        \item[iv.] Notice  that the out-degree, out-closeness, and betweenness centrality measures are sensitive to noise. Particularly, for lower degrees of freedom, the bias of raw data estimates increases. Overall, the results highlight how greatly misleading centrality measures could be based on raw data without proper denoising.
    \end{itemize}
    
        Figure \ref{c2} implies that $\beta$ clearly has a meaningful effect on the methodology's accuracy. It gives more control to weighing information that influences the updated distributions of scale matrices $\bm{\Sigma}_1, \bm{\Sigma}_2$. Thus, it would be reasonable to consider it unknown and estimate it by imposing a prior distribution. \\
        
        The paper now refines this said control of $\beta$. First, it explores the information that the likelihood has on $\beta$. In other words, it imposes a non-informative prior on $\beta$. The strict and technical interpretation of a prior that is non-informative does not seem to have a consensus among various works in literature. We follow the heuristic approach that a non-informative prior should be invariant under monotone transformation. On this assumption, a suitable non-informative prior is proportional to the square root of Fisher Information, typically denoted $I_F(\beta)$. This is known as Jeffrey's prior \cite{jeffrey}. The Fisher information of $\beta$ is the expected value of the second derivative of the natural log of the prior pdf \ref{priors2}:
        \begin{align}
            I_F(\beta) =& -\mathbb{E} \left(\frac{\partial^2}{\partial^2 \beta}ln(\pi).  \right) 
        \end{align}
        Thus, the non-informative prior pdf of $\beta $ is:
        \begin{align}
            \label{jp}
            f(\beta) \propto \sqrt{I_F(\beta)}.
        \end{align}
        Including pdf (\ref{jp}) to the prior pdf (\ref{priors2}), and using algorithm \ref{gibbs2} with full conditional pdf (\ref{beta3}) the results of the methodology is discussed below.
        \begin{figure}[H]
		\fbox{\parbox{\textwidth}{
			\begin{subfigure}[H]{0.5\textwidth}
				\includegraphics[width=\textwidth]{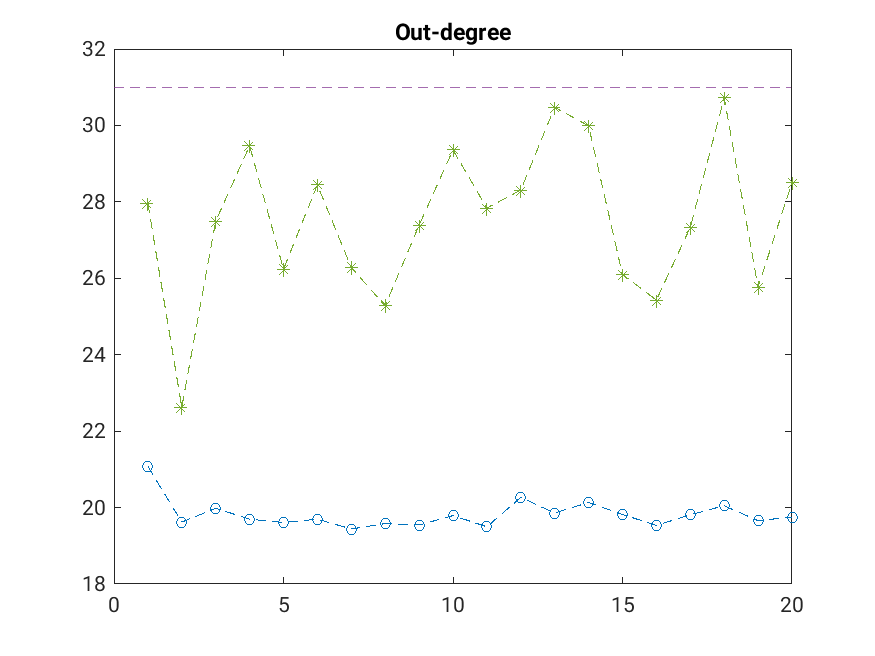}
			\end{subfigure}%
			\begin{subfigure}[H]{0.5\textwidth}
				\includegraphics[width=\textwidth]{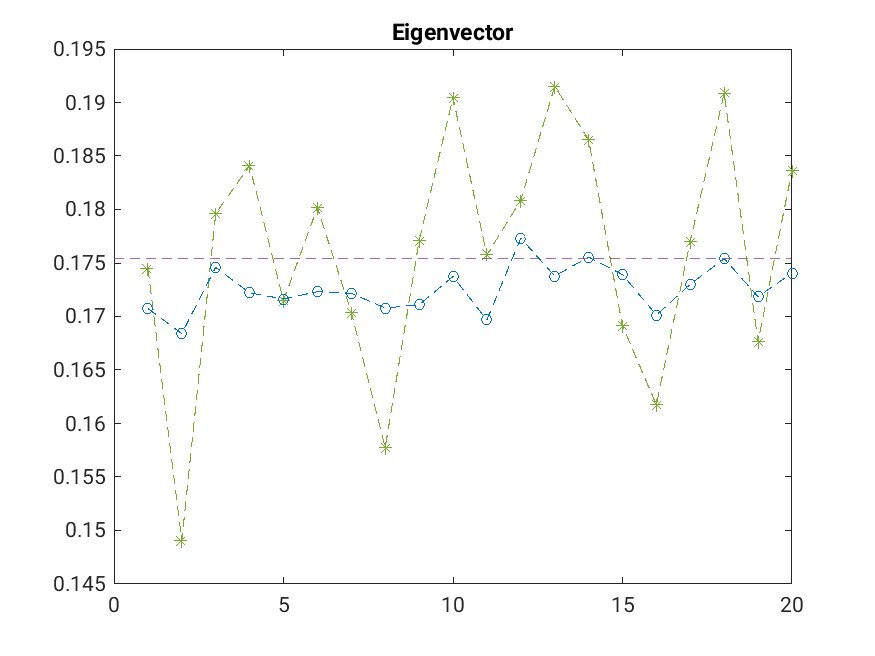}
			\end{subfigure}
			\newline
			\begin{subfigure}[H]{0.5\textwidth}
				\includegraphics[width=\textwidth]{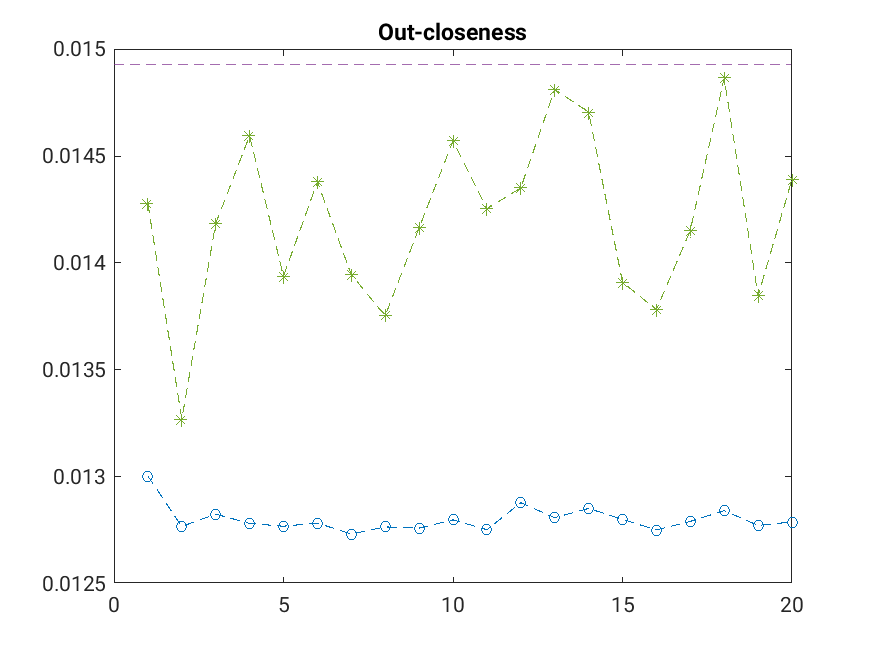}
			\end{subfigure}%
			\begin{subfigure}[H]{0.5\textwidth}
				\includegraphics[width=\textwidth]{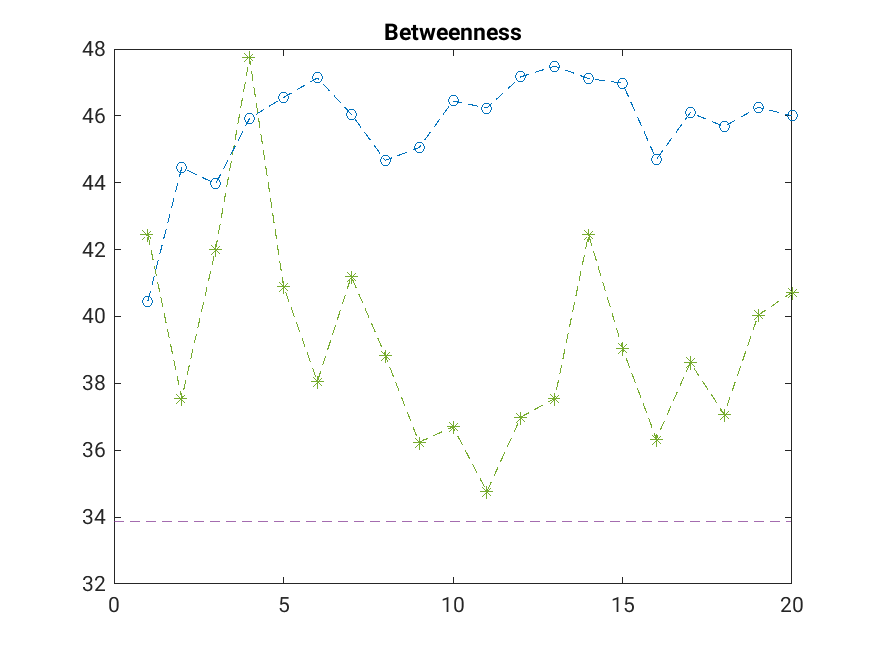}
			\end{subfigure}
		}}
		\caption{The horizontal dashed line represents the true value, the blue bubble-dashed line represents the raw data averages, and the asterisk-dashed lines are the averages of samples obtained from Algorithm \ref{gibbs2} using non-informative prior for $\beta$ (\ref{jp}).}
        \label{c3}
	\end{figure}
    	\begin{figure}[H]
		\centering
		\fbox{\begin{subfigure}[H]{0.5\textwidth}
				\includegraphics[width=\textwidth]{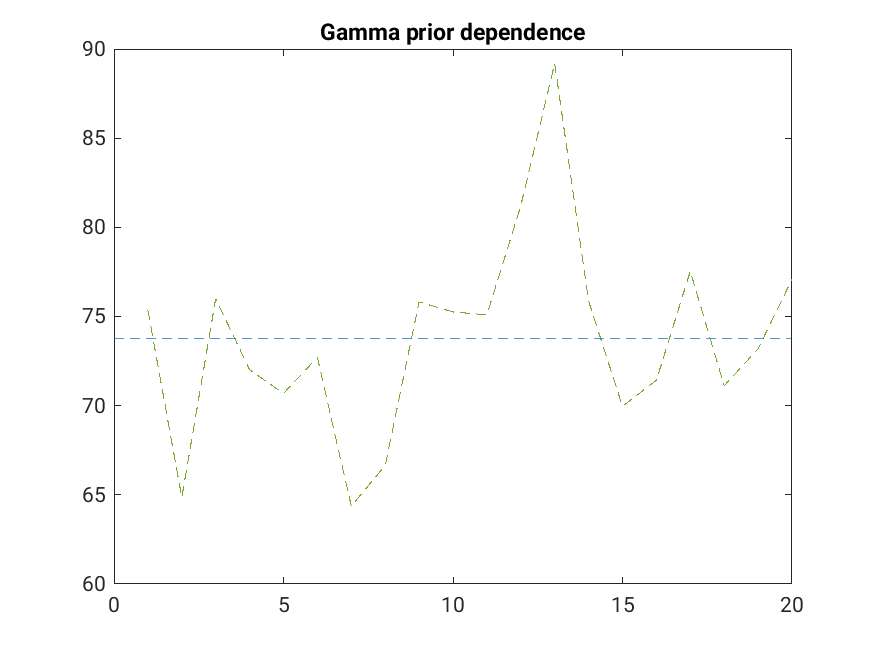}
			\end{subfigure}%
			\begin{subfigure}[H]{0.5\textwidth}
				\includegraphics[width=\textwidth]{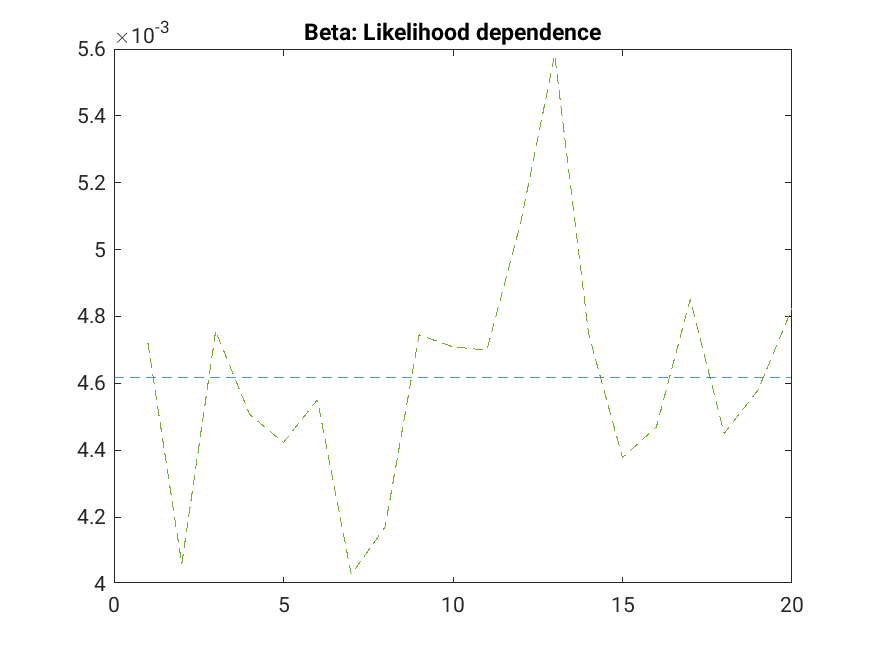}
		\end{subfigure}}
		\caption{Averages of $\gamma$ (left) and $\beta$ (right) for various degrees of freedom.}
		\label{c3gb}
	\end{figure}
    The following is observed from Figures \ref{c3} and \ref{c3gb}:
    \begin{itemize}
        \item[i.] The likelihood does contain information to improve accuracy on the out-degree, out-closeness, and betweenness.
        \item[ii.] In particular, there is an element of flexibility and robustness that a non-informative prior has on $\beta$. If prior information is not readily available, a non-informative prior still provides a great improvement to accuracy compared to raw data averages.
        \item[iii.] There is an apparent direct relationship between the values $\gamma$ and $\beta$ assume. That is, an increase of $\gamma$ occurs when $\beta$ increases and a decrease of $\gamma$ occurs when $\beta$ decreases.
    \end{itemize}
    Note that a non-informative prior did not improve the methodology's performance with regard to estimating a graph's eigencentrality. The same is true for a graph's betweenness with adjacency matrices from a matrix variate t-distribution with degrees of freedom close to 1. Thus, it is worth investigating further the effect of prior knowledge on $\beta$. To that end, the paper imposes an inverse gamma distribution with pdf:
    \begin{align}
        \label{ig}
		f(\beta) = \frac{1}{b_{\beta}^{a_{\beta}} \Gamma(a_{\beta})} \beta^{-(a_{\beta}+1)}e^{(-\frac{1}{b_{\beta}\beta})},
    \end{align}
    where the shape and scale parameters $a_{\beta}, b_{\beta} >0$ respectively.
    Using pdf (\ref{ig}) to prior pdf (\ref{priors2}) and using algorithm \ref{gibbs2} with full conditional pdf (\ref{beta3}), the results of the methodology is given below (see Figure 7).
    	\begin{figure}[H]
		\fbox{\parbox{\textwidth}{
				\begin{subfigure}[H]{0.5\textwidth}
					\includegraphics[width=\textwidth]{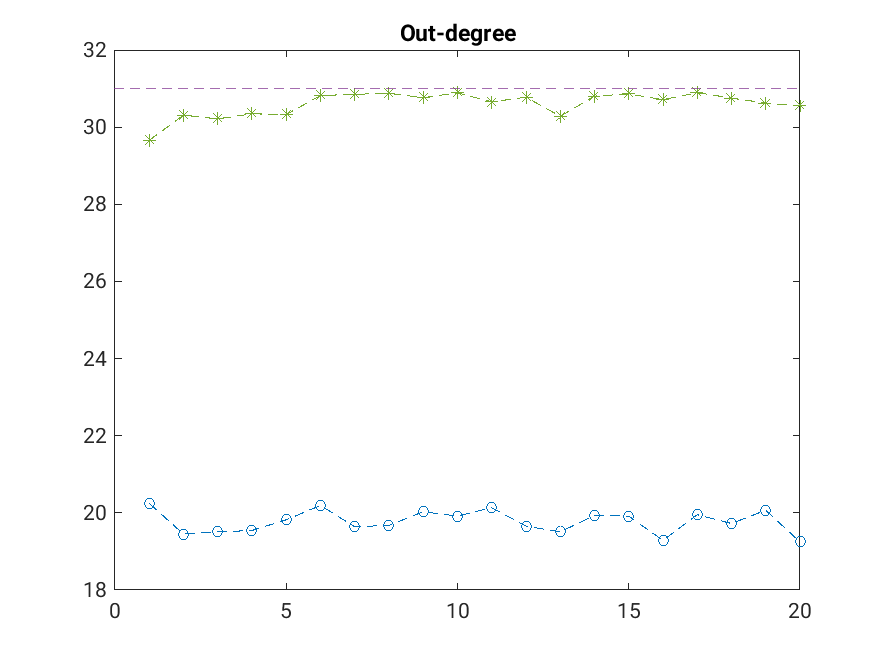}
				\end{subfigure}%
				\begin{subfigure}[H]{0.5\textwidth}
					\includegraphics[width=\textwidth]{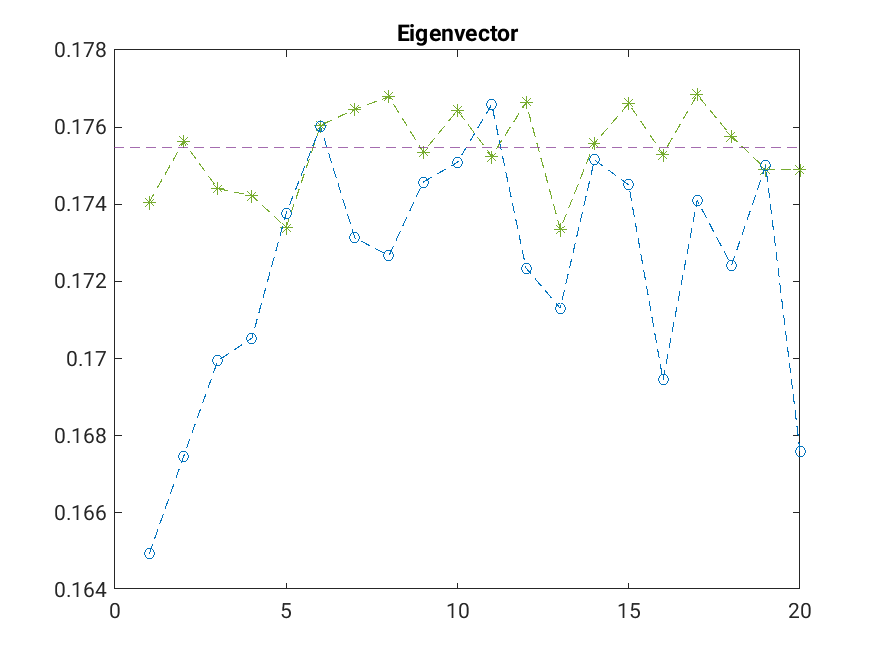}
				\end{subfigure}
				\newline
				\begin{subfigure}[H]{0.5\textwidth}
					\includegraphics[width=\textwidth]{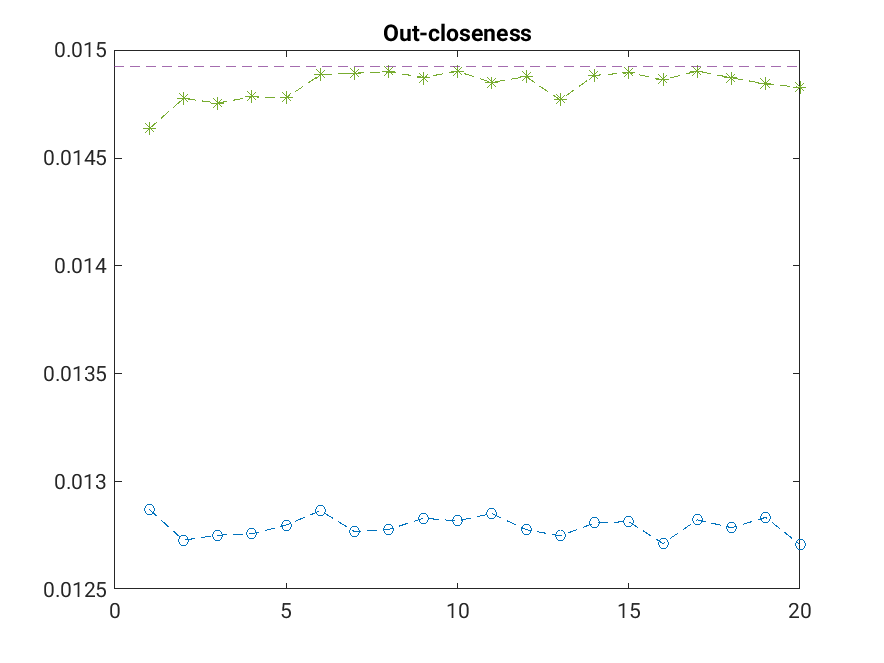}
				\end{subfigure}%
				\begin{subfigure}[H]{0.5\textwidth}
					\includegraphics[width=\textwidth]{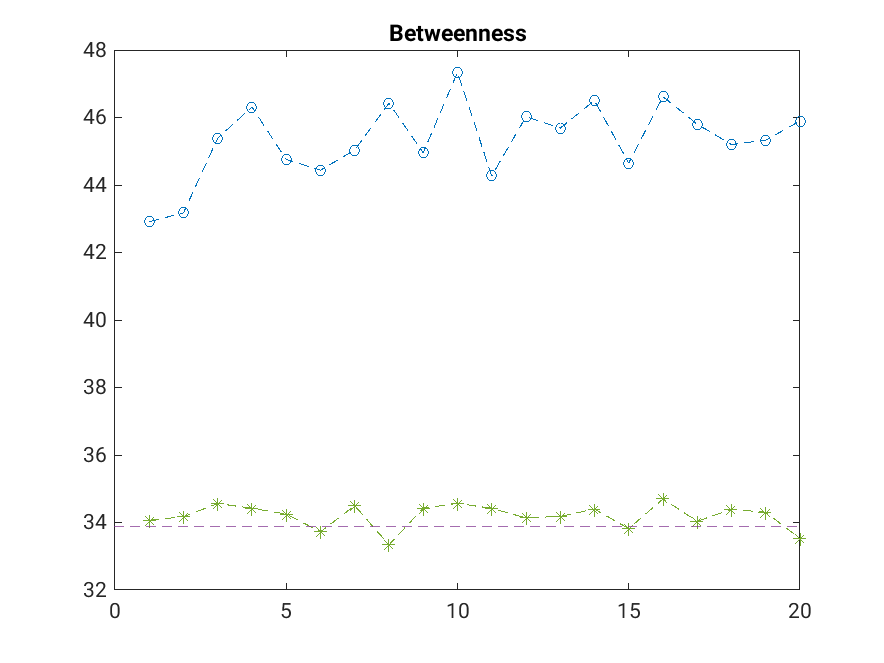}
				\end{subfigure}
		}}
		\caption{Estimation results: The horizontal dashed line represents the true value, the blue bubble-dashed line represents the raw data averages, and the asterisk-dashed lines are the averages of samples obtained from algorithm \ref{gibbs2} using informative prior (\ref{ig}) for $\beta$.}
        \label{informative}
	\end{figure}
	\begin{figure}[H]
		\centering
		\fbox{\begin{subfigure}[H]{0.5\textwidth}
				\includegraphics[width=\textwidth]{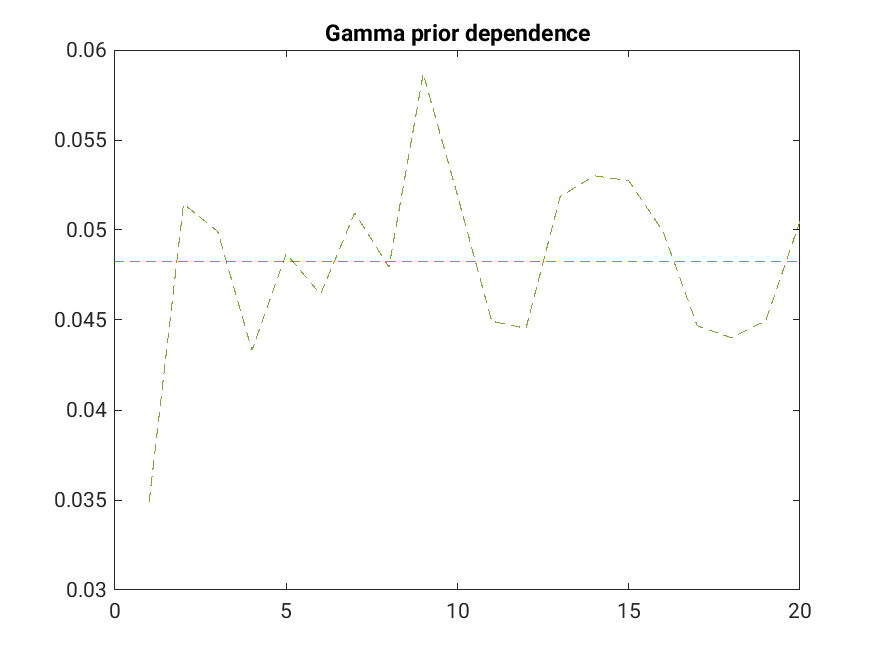}
			\end{subfigure}%
			\begin{subfigure}[H]{0.5\textwidth}
				\includegraphics[width=\textwidth]{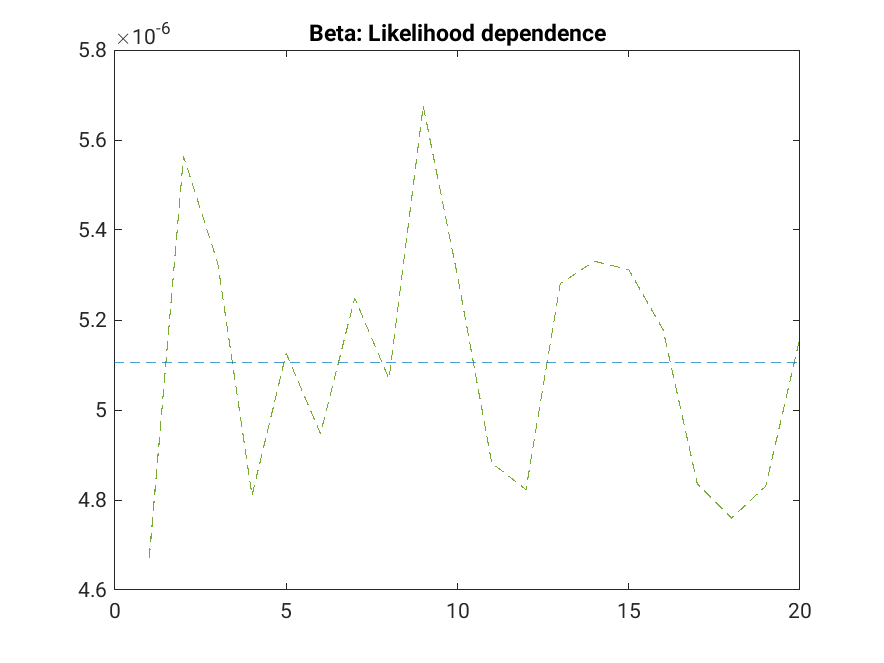}
		\end{subfigure}}
		\caption{Averages of $\gamma$ (left) and $\beta$ (right) for various degrees of freedom.}
        \label{inf_bg}
	\end{figure}
    Refer to Figure \ref{informative}. Note the clear and impressive improvement after imposing an informative prior on $\beta$. Particular attention to the methodology's accuracy with respect to eigencentrality, and the overall performance for observations from a matrix variate distribution with degrees of freedom close to one. Recall that the lower the degrees of freedom, the larger the kurtosis is for the t-distribution. Despite the excess kurtosis, there is a large difference between the raw data averages and that of the proposed methodology. From Figure \ref{inf_bg} there is a large difference in the magnitudes of $\gamma$ and $\beta$ compared to Figure \ref{c3gb}. That is, $\gamma$ does not assume large values to correct the estimation. Hence, an informative prior makes a meaningful difference in how much information it gives to $\beta$.\\
    
    One property to note of the $\mathcal{MG}_n(\delta_1,\beta,\bm{\Phi}_1)$ distribution is that if $\delta_1= \frac{\kappa}{2}$ and $\beta = 2$ then the pdf simplifies to that of a Wishart distribution with $\kappa$ degrees of freedom and scale matrix $\bm{\Phi}_1$. In other words it is $\mathcal{W}_n(\kappa, \bm{\Phi})$ distributed. Similarly, if $\delta_2 = \frac{\kappa}{2}$ and $\beta = 2$ then the pdf simplifies to that of an $\mathcal{IW}_n(\kappa, \bm{\Phi}_2)$ distribution. Thus, we compare the performance of the methodology under the special case $\beta=2$ versus the case $\beta$ is estimated. That is, we explore the added benefit that $\beta$ has on improving the accuracy of the methodology. \\
     \\
     Refer to error plots in Figure \ref{comparisons} below: Notice that the error points for $\beta$ estimated are closer to the zero reference line among all centrality measures. In other words, the errors are consistently smaller across all centrality measures and for various degrees of freedom when $\beta$ is estimated compared to assuming the special case that $\beta=2$. An especially crucial highlight is an impressive difference in errors for degrees of freedom close to 1. In the presence of increased kurtosis, this extra parameter $\beta$ manages to improve estimation accuracy.
	\begin{figure}[H]
		\fbox{\parbox{\textwidth}{
				\begin{subfigure}[H]{0.5\textwidth}
					\includegraphics[width=\textwidth]{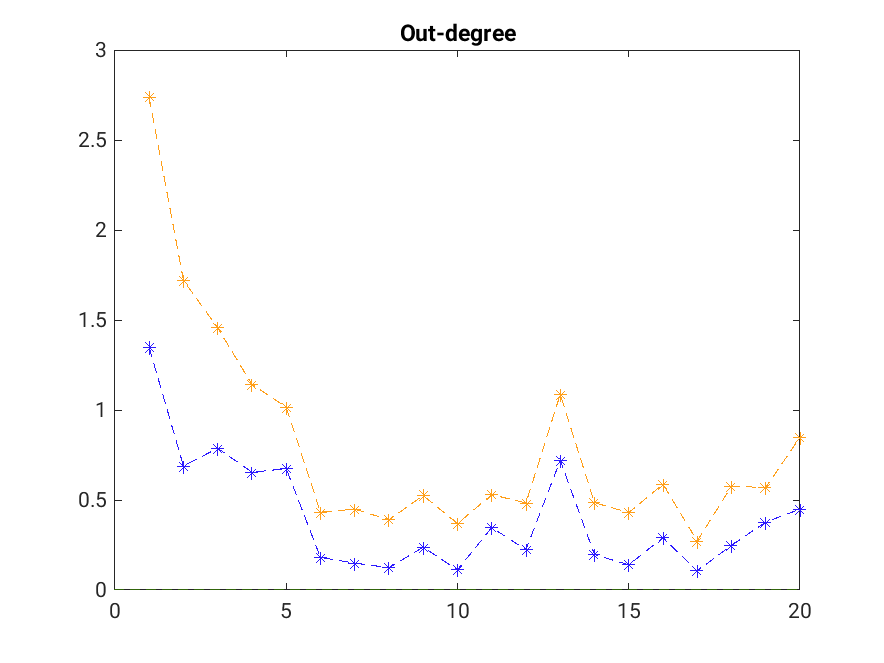}
				\end{subfigure}%
				\begin{subfigure}[H]{0.5\textwidth}
					\includegraphics[width=\textwidth]{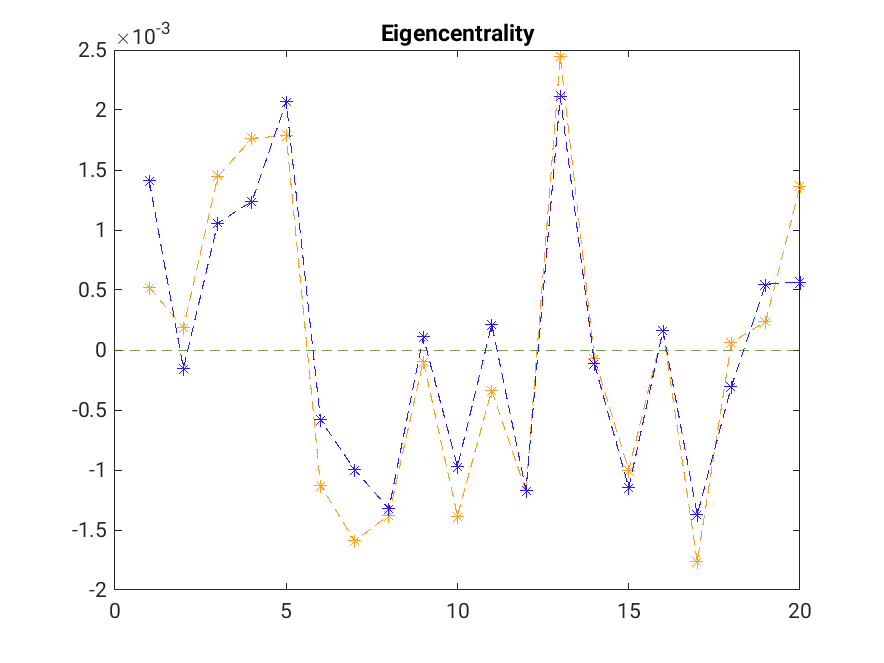}
				\end{subfigure}
				\newline
				\begin{subfigure}[H]{0.5\textwidth}
					\includegraphics[width=\textwidth]{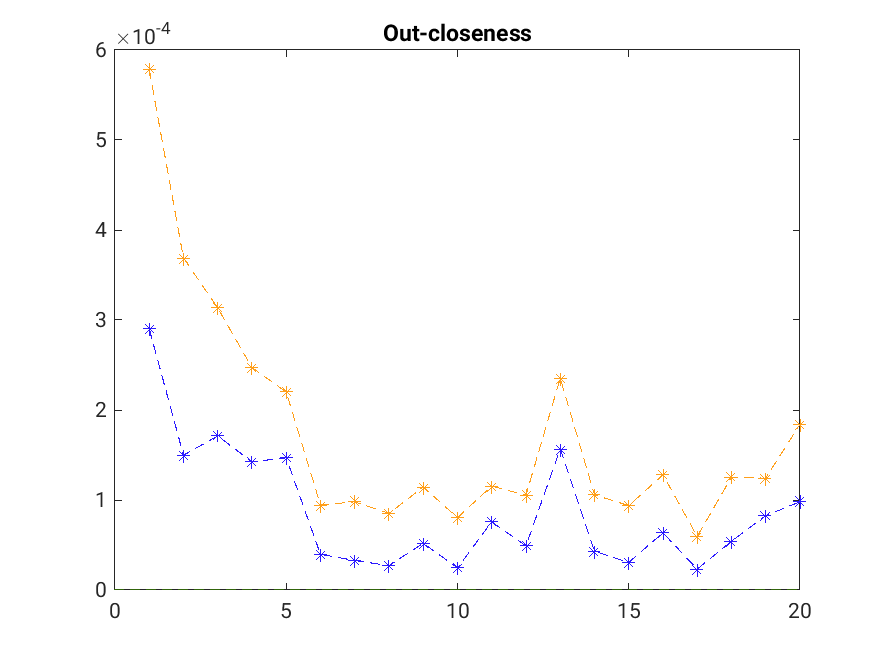}
				\end{subfigure}%
				\begin{subfigure}[H]{0.5\textwidth}
					\includegraphics[width=\textwidth]{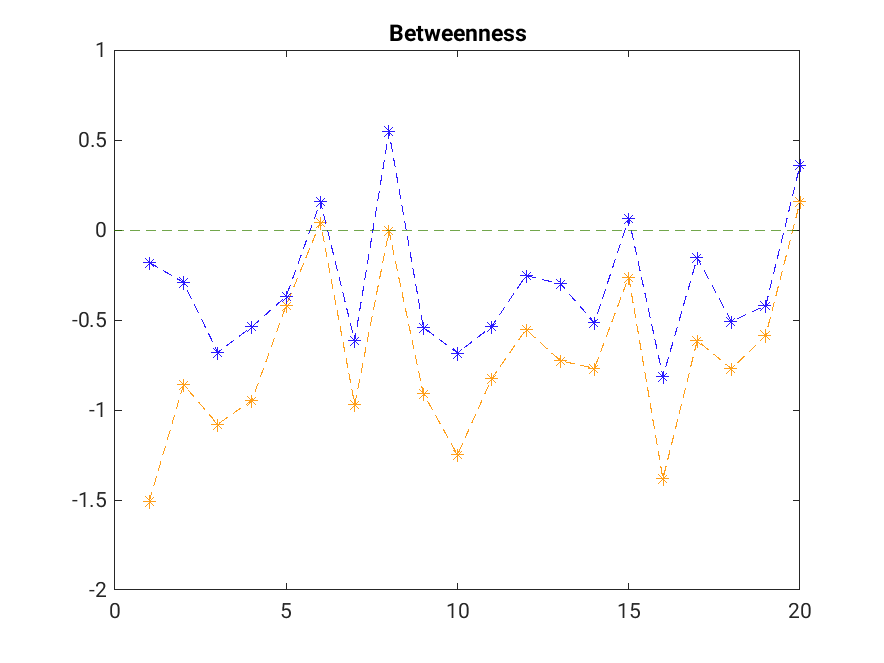}
				\end{subfigure}
		}}
		\caption{Estimation error plots: Orange and blue lines represent errors when $\beta=2$ and when $\beta$ is estimated using informative prior pdf \ref{ig}, respectively.}
  \label{comparisons}
	\end{figure}
	\subsection{Data application}
 The methodology is applied to the weekly stock prices of 70 European firms from 4 January 2016 to 31 December 2019, excluding the range of time COVID was present, resulting in 105 observations per firm. The data comprises of a range of Global Industry Classification Standard sectors (or GICS sectors), namely: Communication Services (5), Consumer Discretionary (15), Consumer Staples(6), Energy (2), Financials (11), Health Care (6), Industrials (10), Information Technology (5), Materials (2), Food and Beverages (1), Real Estate (3), and Utilities (5). The firms also range over various countries in Europe: Germany (28), France (37), and Italy (5). Granger causality hypothesis tests are applied pairwise for week $t$. The resulting test statistics belong in a matrix that is an observed $\bm{Y}_t$.\\

We employ the well-known centrality measures as was done in the simulation experiments. Comparing how similar or different the scores computed between raw data estimates and that of the proposed methodology can effectively showcase the significance of the latter. 
	\begin{figure}[h]
		\fbox{\parbox{\textwidth}{
			\begin{subfigure}{0.5\textwidth}
				\includegraphics[width=\textwidth]{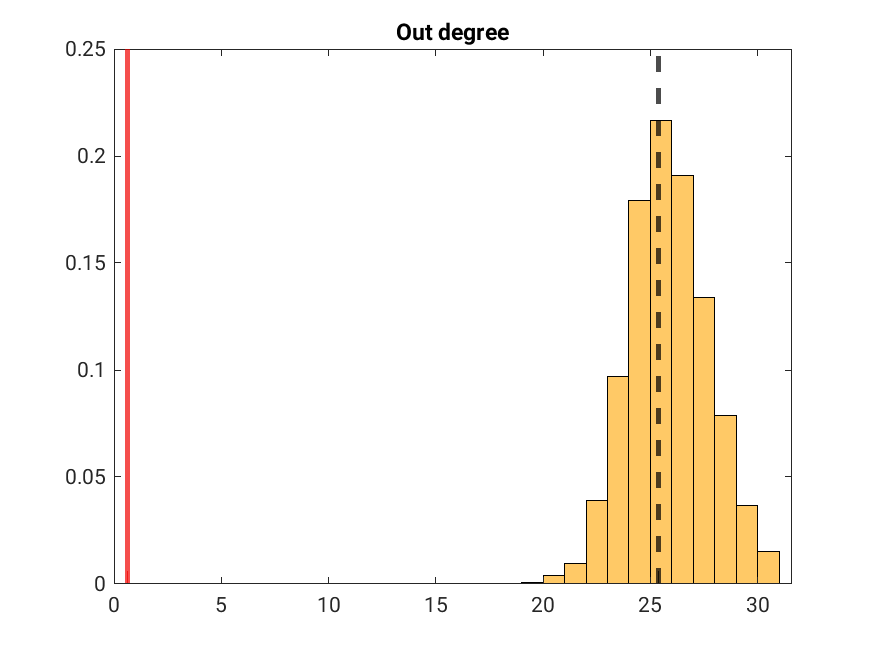}
			\end{subfigure}%
			\begin{subfigure}{0.5\textwidth}
				\includegraphics[width=\textwidth]{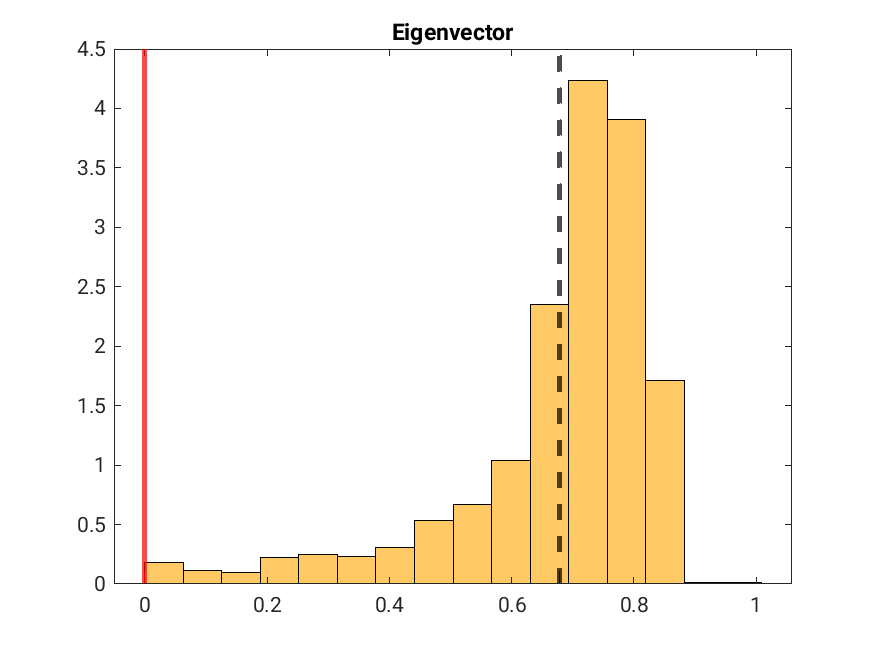}
			\end{subfigure}
   \newline
			\begin{subfigure}{0.5\textwidth}
				\includegraphics[width=\textwidth]{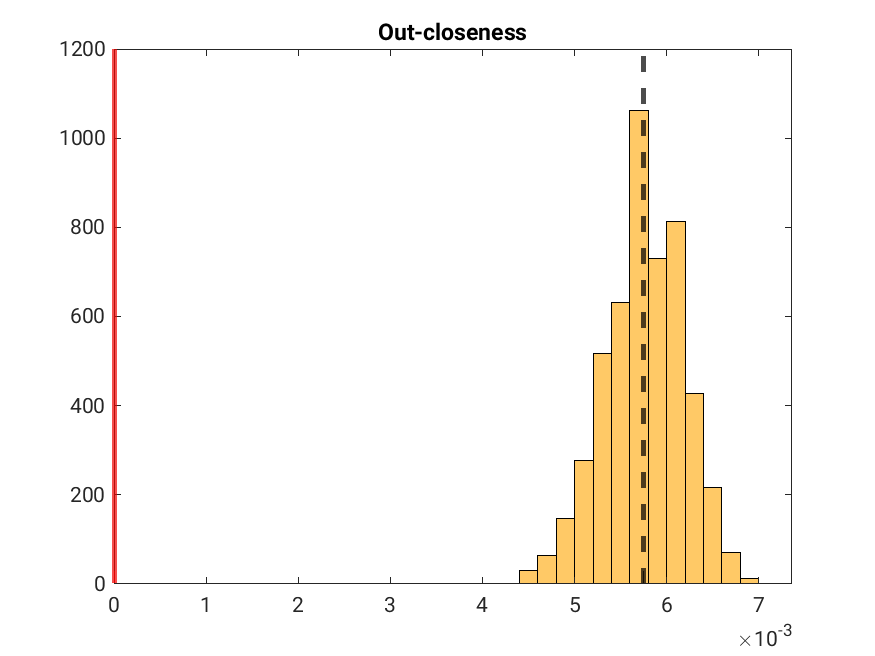}
			\end{subfigure}%
			\begin{subfigure}{0.5\textwidth}
				\includegraphics[width=\textwidth]{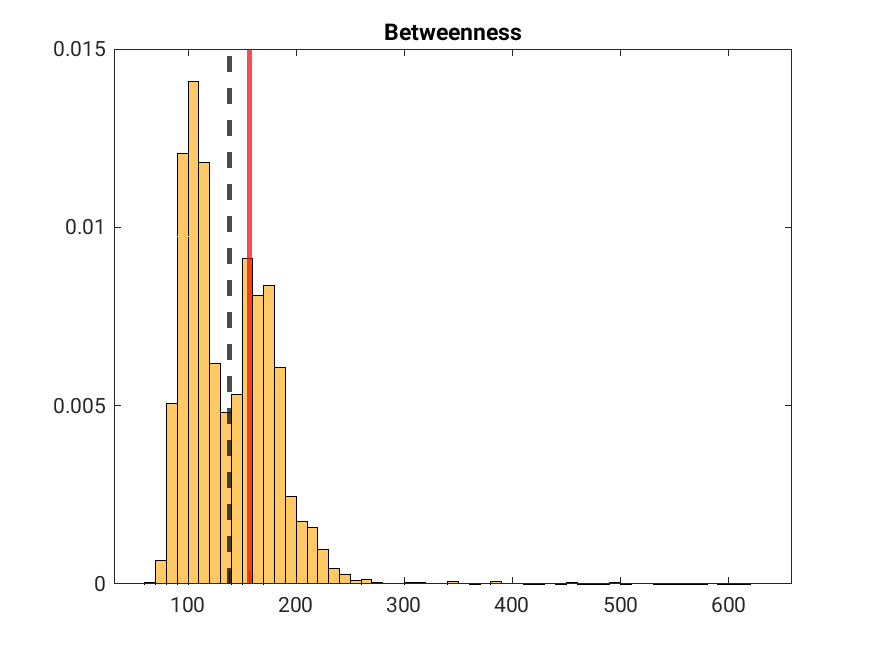}
			\end{subfigure}
		}}
		\caption{Estimated centrality measures: The solid red line and dashed black lines represent the averages of the raw data, and methodology respectively.}
		\label{fig:1}		
	\end{figure}
	The results from the application are shown in Figure \ref{fig:1}. It is observed that there are clear discrepancies between the different estimators, with particular attention to the out-degree, out-closeness, and eigencentrality. The raw data seems to underestimate the centrality measures. In other words, the discrepancies highlight how noise left in data induces heavy bias in the estimation procedure. If not dealt with, the noise may jeopardise the validity and reliability of analysis built on data.\\

\begin{figure}[H]
		\fbox{\parbox{\textwidth}{
			\begin{subfigure}{0.5\textwidth}
				\includegraphics[width=\textwidth]{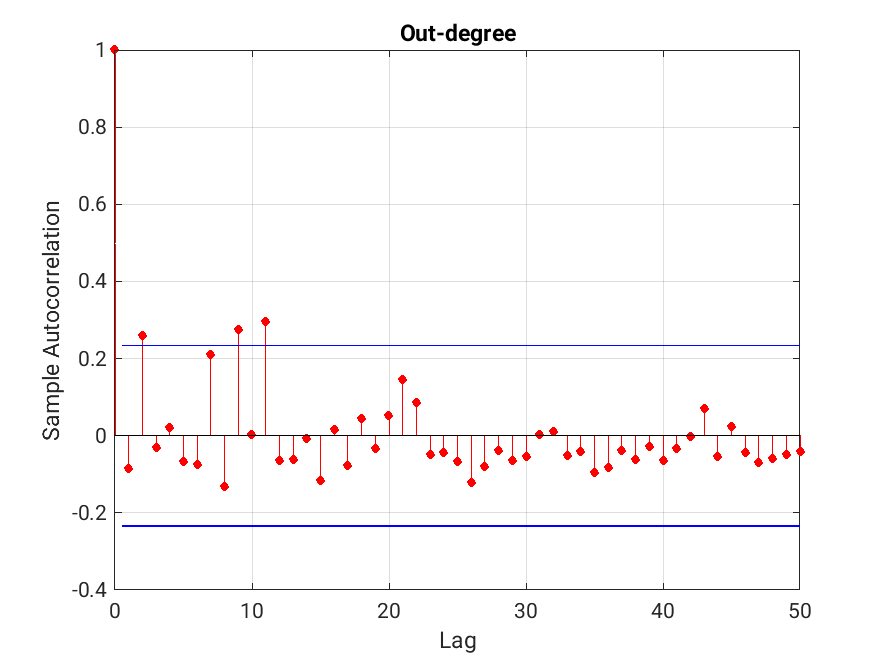}
			\end{subfigure}%
			\begin{subfigure}{0.5\textwidth}
				\includegraphics[width=\textwidth]{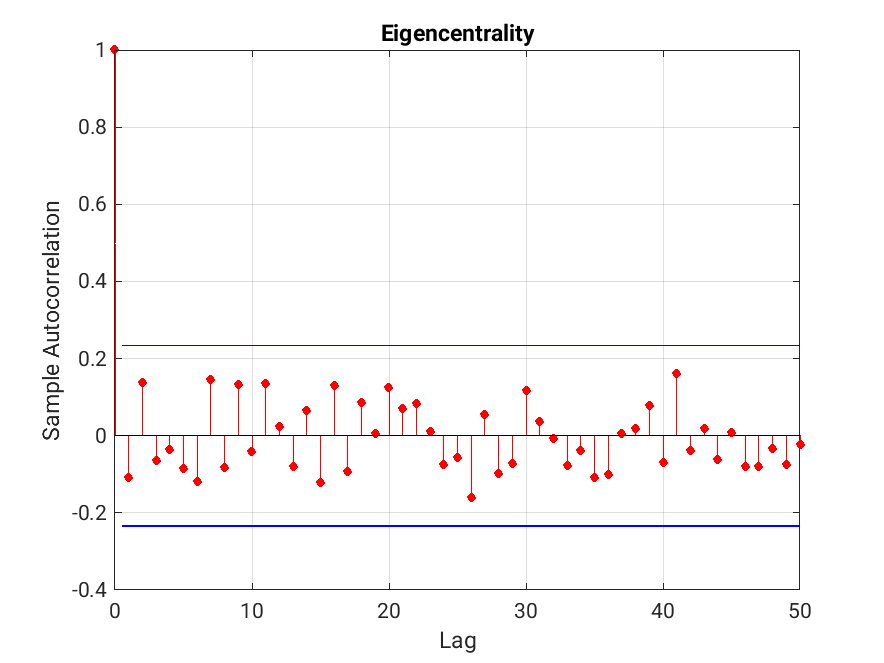}
			\end{subfigure}
   \newline
			\begin{subfigure}{0.5\textwidth}
				\includegraphics[width=\textwidth]{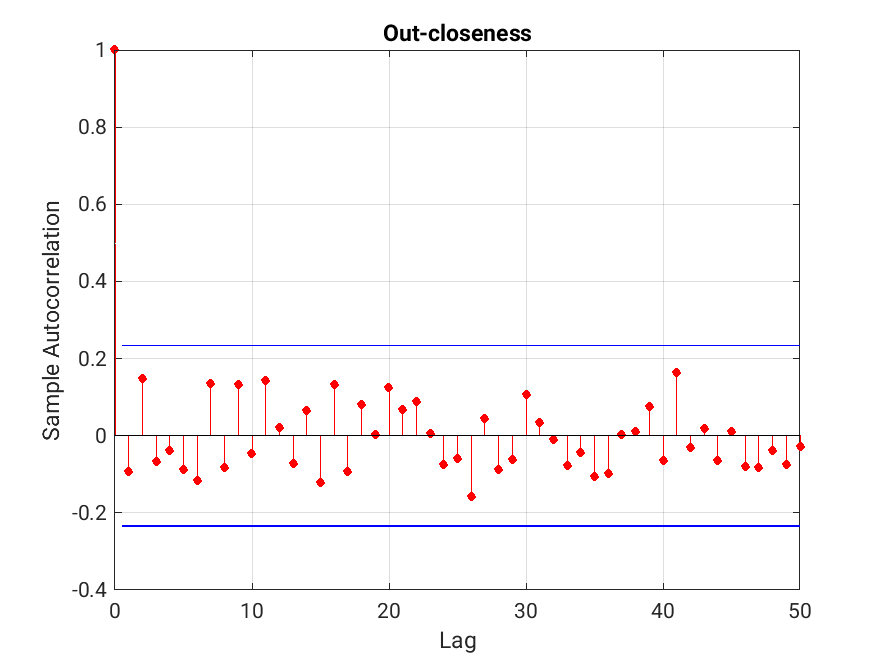}
			\end{subfigure}%
			\begin{subfigure}{0.5\textwidth}
				\includegraphics[width=\textwidth]{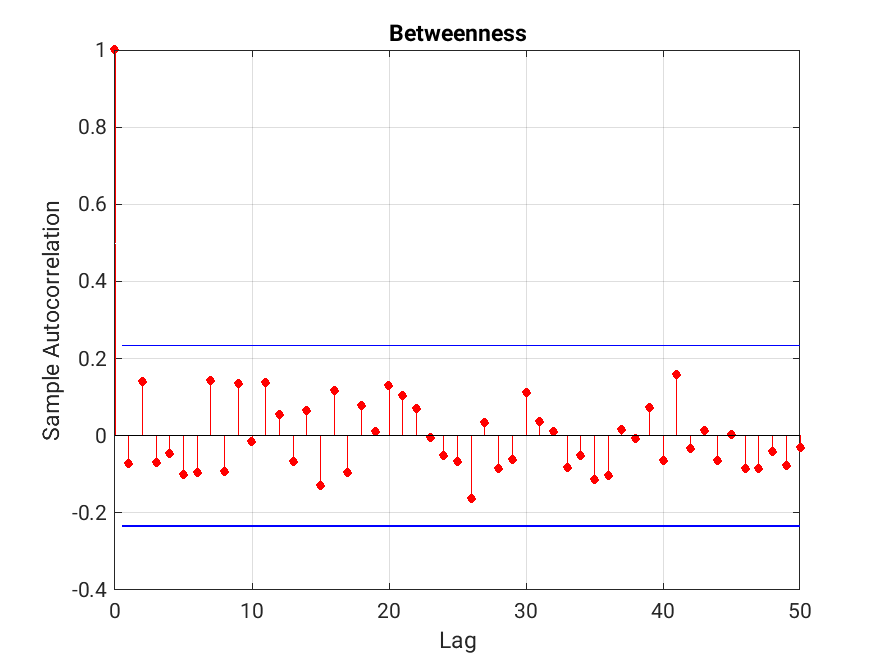}
			\end{subfigure}
		}}
		\caption{Convergence check of Gibbs sampler: Autocorrelation plots of averaged centrality measures. The red pointed lines are the sample autocorrelation for each lag and the horizontal solid blue lines are the 95\% confidence intervals for insignificant autocorrelation.}
		\label{convergence}		
	\end{figure}

    The sample obtained from algorithm \ref{gibbs2} has autocorrelations that are insignificant on a 5\% level, see Figure \ref{convergence}. Thus, it is reasonable to assume convergence of the algorithm \cite{gibbs_sampler}.
    
	\section{Conclusion}
This article introduces for the first time a flexible prior for the covariance structure of a matrix variate t for the noise process, following a Gaussian graphical model. A simulation experiment demonstrated empirically  the results of the methodology in a way that is representative of the methodology’s application to real-life stock data on 70 European firms. The results from the simulations and empirical experiments highlight the added flexibility and control that a generalised prior has on the methodology and demonstrates that our proposed framework outperforms the previously considered work. The importance of flexible distributions are emphasized amongst others, by \cite{ley2015flexible}, \cite{jones2015families} and \cite{ley2021flexible}.  The effect of prior information to better accommodate the nature of data on modelling variance robustly and flexibly has been overlooked in literature.  While this paper focused on denoising financial data, it extended the framework of denoising data to any field where applicable. It described the assumptions about the nature of errors in data and thus established their characterisation and probabilistic nature. It extensively investigated the Bayesian approach to estimate the noisy connectivity structure.  A range of centrality measures is applied to the estimated network for a comprehensive summary of the estimation’s accuracy and robustness.  In particular, it adds to the theoretical toolkit researchers may need for more robust and accurate network estimation. Future studies can extend the denoising method proposed in this paper to consider a generalized scalar Wishart distribution or the generalized diagonal Wishart distribution of \cite{adhikari2010generalized}.

\section*{Acknowledgements}
This work was based upon research supported in part by the National Research Foundation (NRF) of South Africa (SA), grant RA201125576565 \&  RA211204653274, nr 145681 \& 151035; NRF ref. SRUG2204203865, the Centre of Excellence in Mathematical and Statistical Sciences, based at the University of the Witwatersrand (SA). The opinions expressed and conclusions arrived at are those of the authors and are not necessarily to be attributed to the NRF. Mohammad Arashi's work is based on the research supported in part by the Iran National
Science Foundation (INSF) grant No. 4015320. The authors would like to thank Professor Monica Billio, University of Venice, for providing the data  used in this paper.
\appendix

\section{Definitions of goodness-of-fit measures}
Centrality is a comprehensive way to evaluate the goodness-of-fit of the proposed estimation techniques. No one measure is not a one-size-fits-all, as the interpretation of centrality varies greatly depending on the context of the field of study. Here the following measures are reviewed that cover a broad perspective of node influence that can be used to evaluate the accuracy of the methodology discussed. For all definitions below \cite{sna}, consider a graph with corresponding adjacency matrix $\bm{A}$. Let the $ij^{th}$ entry of $\bm{A}$ be denoted $a_{i,j}$.
	\begin{definition}
		Consider the number of edges that leave a node $i$. The more edges that leave $i$, the more connected it is to other nodes and thus more central in the graph. This is known as a node's out-degree. Mathematically let,
		\begin{equation}
			\label{degree}
			d^{(i)} = \sum_{j=1, j\ne i}^{n} a_{i,j}.
		\end{equation}
		Then $d^{(i)}$ is the out-degree of node $i$.
		\newline
	\end{definition}
	\begin{definition}
		\label{closeness}
		Consider path lengths from node $i$ to nodes $j=1,\dots,n, j \ne i$. Heuristically, it is easy to see that a node with shorter lengths to other nodes implies it is easier for said node to reach other nodes. This is known as a node's out-closeness. Mathematically let, 
		\begin{equation}
			c^{(i)} = \left\{\sum_{j=1, j\ne i}^{n} l(i,j)\right\}^{-1}.
		\end{equation}
		Then $c^{(i)}$ is the out-closeness of node $i$. Note that there is a normalised variation of $c^{(i)}$ given as
		\begin{equation}
			\label{ncloseness}
			c^{(i)}_{norm} = \frac{R_i}{(n-1)} \left\{\sum_{j=1, j\ne i}^{n} l(i,j)\right\}^{-1},
		\end{equation}
		where $R_i$ is the number of reachable nodes from node $i$.
		$c^{(i)}_{norm}$ can be interpreted as the averaged shortest path from node $i$. The normalised variant is useful to compare the closeness of nodes across graphs of different sizes and account for unreachable nodes.
	\end{definition}
	\begin{definition}
		Let $n(i,j)$ denote the number of paths from node $i$ to node $j$ and let $n_v(i,j)$ denote the number of paths from $i$ to $j$ that pass through node $v$. Then it is possible to measure the transmittable influence of node $v$ as the proportion number of paths from $i$ to $j$ that passes through $v$. Clearly, a large proportion indicates that node $v$ acts as an influential transmitter between nodes. Mathematically let,
		\begin{equation}
			\label{betweenness}
			b^{(v)} = \sum_{i=1,v \notin\{i,j\}}^{n}\sum_{i=1, j\ne i}^{n} \frac{n_v(i,j)/n(i,j)}{(n-1)(n-2)}.
		\end{equation}
		Then $b^{(v)}$ is the betweenness of node $v$ where $(n-1)(n-2)$ represents the total number of node pairs to sum through.
	\end{definition}
	\begin{definition}
		All definitions above measure the centrality of a node without taking into account the centrality scores of other nodes. It might be useful to measure how influential a node is relative to others. Hence, denote $x^{(i)}$ the relative influential score of node $i$ and $\lambda$ the largest eigenvalue of $\bm{A}$. Then,
		\begin{equation}
			\label{evc}
			x^{(i)} = \frac{1}{\lambda} \sum_{j=1}^{n}a_{i,j}x^{(j)},
		\end{equation} 
		is known as the eigenvector centrality of node $i$.
	\end{definition}

 \bibliographystyle{elsarticle-num}





% \end{thebibliography}
\end{document}